\documentclass[11pt,twoside,a4paper,cmspaper,final,collab]{cms-tdr}
\def\svnVersion{d199d06}\def\svnDate{2026/05/27}\def\cmsCernNoTag{CERN-EP-2025-238}\def\cmsCernDate{\today}\def\cmsMessage{Published in the Journal of High Energy Physics as \href{http://dx.doi.org/10.1007/JHEP05(2026)191}{\doi{10.1007/JHEP05(2026)191}.}}
\begin{document}\cmsNoteHeader{EXO-24-033}

\newcommand{\dxy}{\ensuremath{d_{xy}}\xspace}
\newcommand{\dz}{\ensuremath{d_{z}}\xspace}
\newcommand{\dm}{\ensuremath{\Delta m}\xspace}
\newcommand{\dxysig}{\ensuremath{S_{xy}^{\text{trk}}}\xspace}
\newcommand{\lxy}{\ensuremath{L_{xy}}\xspace}
\newcommand{\lxysig}{\ensuremath{S_{xy}^{\text{vtx}}}\xspace}
\newcommand{\ctau}{\ensuremath{c\tau}\xspace}
\newcommand{\mstop}{\ensuremath{m_{\PSQt}}\xspace}
\newcommand{\dphi}{\ensuremath{\Delta \phi}\xspace}
\newcommand{\deta}{\ensuremath{\Delta \eta}\xspace}
\newcommand{\br}{\ensuremath{\mathcal{B}(\PSQt \to \PQb \Pf\overline{\Pf}'\PSGczDo)}\xspace}
\newcommand{\mllp}{\ensuremath{m_{\text{LLP}}}\xspace}
\newcommand{\alphap}{\ensuremath{\alpha_{\text{p}}}\xspace}

\cmsNoteHeader{EXO-24-033}
\title{Search for long-lived particles using displaced vertices with low-momentum tracks in proton-proton collisions at \texorpdfstring{$\sqrt{s} = 13\TeV$}{sqrt(s) = 13 TeV}}

\date{\today}

\abstract{A search for long-lived particles using final states including a displaced vertex with low-momentum tracks, large missing transverse momentum, and a jet from initial-state radiation is presented. This search uses proton-proton collision data at a center-of-mass energy of 13\TeV collected by the CMS experiment at the CERN LHC in 2017 and 2018, with a total integrated luminosity of 100\fbinv. This analysis adopts specific supersymmetric (SUSY) coannihilation scenarios as benchmark signal models, characterized by a next-to-lightest SUSY particle (NLSP) with a mass difference of less than 25\GeV relative to the lightest SUSY particle, assumed to be a bino-like neutralino. In the top squark (\PSQt) NLSP model, the NLSP is a long-lived \PSQt, while in the bino-wino NLSP scenario, the mass-degenerate NLSPs are a  wino-like long-lived neutralino and a short-lived chargino. The search excludes top squarks with masses less than 400--1100\GeV and wino-like neutralinos with masses less than 220--550\GeV, depending on the signal parameters, including the mass difference, mass, and lifetime of the long-lived particle. It sets the most stringent limits to date for the \PSQt and bino-wino NLSP models.}

\hypersetup{
pdfauthor={CMS Collaboration},
pdftitle={Search for long-lived particles using displaced vertices with low-momentum tracks and missing transverse momentum in proton-proton collisions at sqrt(s)=13 TeV},
pdfsubject={CMS},
pdfkeywords={CMS, displaced vertex, ISR jet}} 

\maketitle 

\section{Introduction}

Many theories beyond the standard model (SM) predict long-lived particles (LLPs), whose mean proper lifetimes are longer than ${\sim}0.1\unit{ps}$, which could be detected by the experiments at the CERN LHC. In some of these models, LLP decays include an invisible particle, which can be a dark-matter candidate, making those models even more appealing. Such models include the top squark (\PSQt) next-to-lightest supersymmetric particle (NLSP)~\cite{Delgado:2012eu,Grober:2014aha} and bino-wino NLSP~\cite{Nagata:2015pra} models. Additionally, a variety of other models predict similar final states, such as split supersymmetry (SUSY)~\cite{Giudice:2004tc, Hewett:2004nw,ArkaniHamed:2004yi,Gambino:2005eh,Arvanitaki:2012ps,ArkaniHamed:2012gw}, stealth SUSY~\cite{Fan:2011yu,Fan:2012jf}, gauge-mediated SUSY breaking~\cite{GIUDICE1999419,Meade:2008wd,Buican:2008ws}, and hidden valley models~\cite{Strassler:2006im,Strassler:2006ri,Han:2007ae}. Because of their long lifetimes, LLPs can travel a measurable distance in the detector, thus creating unique signatures that enable the possibility to explore new regions of phase space.

This search targets LLPs that propagate between approximately 0.01 and 40\cm before decaying to at least one neutral particle and a number of charged particles. The charged particles from the LLP decays originate from displaced vertices, which can be reconstructed as the common origin point of the trajectories of the charged LLP decay products. In addition, the invisible particles produce missing transverse momentum (\ptmiss), while an initial-state radiation (ISR) jet provides the recoil against the LLPs to yield a large \ptmiss. As a result, the search signature includes at least one displaced vertex, \ptmiss, and a jet from ISR.

The ATLAS and CMS Collaborations have previously performed searches for LLPs using displaced vertices~\cite{ATLAS:2017tny, ATLAS:2020xyo, EXO-19-013, CMS:2020iwv, ATLAS:2023oti,CMS:2024trg,CMS:2024xzb}. However, event and vertex selections applied in those searches resulted in a reduced sensitivity to scenarios where the mass difference between the LLP and the invisible particle is less than 25\GeV. The search presented in this paper aims to cover this gap, focusing on a mass difference in the range of 12--25\GeV. To enhance the sensitivity to small mass differences, this search makes use of well-measured low-momentum tracks. Because of the improved performance of the upgraded tracker in 2017--2018 and the reduced displaced tracking efficiency resulting from the electronic saturation effect of the tracker strip readout in 2016~\cite{Cardwell:2021lco}, we perform the search using only events collected in 2017 and 2018, corresponding to an integrated luminosity of 100\fbinv.

\begin{figure}[htbp]
    \centering
    \includegraphics[width=0.49\textwidth]{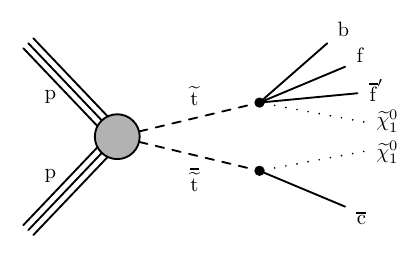}
    \includegraphics[width=0.49\textwidth]{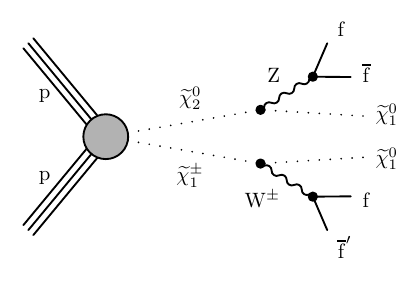}
    \caption{Feynman diagrams for the \PSQt NLSP (left) and the bino-wino NLSP (right) production.}
    \label{fig:diagrams}
\end{figure}

The \PSQt~\cite{Delgado:2012eu,Grober:2014aha} and bino-wino~\cite{Nagata:2015pra} NLSP models, illustrated in Fig.~\ref{fig:diagrams}, are used as benchmark signal models in this search. In both models, a bino-like neutralino (\PSGczDo) is the lightest SUSY particle (LSP). The \PSQt NLSP model describes a coannihilation scenario with a \PSQt as the NLSP. In this model, the \PSQt is pair produced and long lived, and can decay either via a four-body channel into two fermions (\Pf), a \PQb quark, and a \PSGczDo, or via a two-body channel into a \PQc quark and a \PSGczDo. According to the theoretical predictions, the \PSQt mass ($m_{\PSQt}$), the mass difference between the NLSP and LSP (\dm), and the branching fraction of the four-body decay \br can be treated as free parameters, while the \PSQt proper decay length (\ctau) depends on the other parameters. The decay width of the four-body decay $\Gamma(\PSQt \to \PQb \Pf\overline{\Pf}' \PSGczDo)$ is given by Ref.~\cite{Delgado:2012eu} as:
\begin{equation}
\label{eq:stop_br}
\Gamma(\PSQt \to \PQb \Pf\overline{\Pf}'\PSGczDo) = 252\unit{cm}^{-1} \left( \frac{\dm}{30\GeV} \right)^8 \left( \frac{400\GeV}{m_{\PSQt}} \right),
\end{equation}
with $\ctau = \mathcal{B}(\PSQt \to \PQb \Pf\overline{\Pf}'\PSGczDo)/\Gamma(\PSQt \to \PQb \Pf\overline{\Pf}'\PSGczDo)$.
For different fermion flavors in the four-body decay, the corresponding decay widths follow the relationships from Ref.~\cite{Delgado:2012eu}, where \Pell represents a charged lepton (\Pe, \PGm, or \PGt):
\begin{equation}
\label{eq:stop_Wbr}
\Gamma(\PSQt \to \PQb \PQu\overline{\PQd} \PSGczDo) \approx \Gamma(\PSQt \to \PQb \PQc \overline{\PQs} \PSGczDo) \approx 3\Gamma(\PSQt \to \PQb \Pellp \nu_\Pell \PSGczDo).
\end{equation}

When \dm is less than approximately 25\GeV, \ctau becomes sufficiently long to lead to the production of displaced vertices. By requiring one or more displaced vertices, the analysis is sensitive to the scenarios in which the top squarks are long-lived.

The bino-wino NLSP model predicts a nearly mass-degenerate wino-like chargino \PSGcpmDo and a neutralino \PSGczDt as the NLSPs. In this scenario, the \PSGcpmDo decays promptly to a \PWpm~boson and a \PSGczDo, since the decay happens via a dimension-five operator. When the \dm between the NLSP and LSP is less than the \PZ~boson mass, the \PSGczDt predominantly decays through an off-shell \PZ~or \PH~boson, producing a pair of fermions and a \PSGczDo. Both decay channels are suppressed because of the large higgsino mass parameter, rendering the \PSGczDt long-lived. In this search, only \PSGczDt decays mediated by the \PZ~boson are considered. Pair production of \PSGczDt is not considered, since in the pure wino scenario, neutralinos do not couple to the SM \PZ~boson at tree level~\cite{Djouadi:2001kba}. This suppresses the neutral current production mechanism for \PSGczDt pairs.
For the statistical interpretation, the masses of \PSGcpmDo and \PSGczDt, the \dm between the NLSP and LSP, and the \ctau of the \PSGczDt are all treated as free parameters. Previous searches by the ATLAS and CMS Collaborations~\cite{ATLAS:2024qxh,CMS:2024gyw} have explored this model. However, by exploiting displaced signatures, the present search can probe previously inaccessible regions of the parameter space.

The paper is organized as follows. Section~\ref{sec:detector} provides an overview of the CMS detector. Section~\ref{sec:datasets} describes the data, simulation, and event selections used in this search. The vertex reconstruction and selection are discussed in Section~\ref{sec:vtxreco}. Section~\ref{sec:bkgest} presents the definitions of the signal regions (SRs) and the background estimation procedure. Section~\ref{sec:syst} reviews the systematic uncertainties associated with the signal efficiency and the background estimate. The results and statistical interpretation are discussed in Section~\ref{sec:results}. Finally, Section~\ref{sec:summary} summarizes the paper. Tabulated results are provided in the HEPData record for this analysis~\cite{hepdata}.

\section{The CMS detector}
\label{sec:detector}

The CMS apparatus~\cite{CMS:2008xjf,CMS:2023gfb} is a multipurpose, nearly hermetic detector, designed to trigger on~\cite{CMS:2016ngn,CMS:2020cmk,CMS:2024aqx} and identify electrons, muons, photons, and (charged and neutral) hadrons~\cite{CMS:2020uim,CMS:2018rym,CMS:2014pgm}. Its central feature is a superconducting solenoid of 6\unit{m} internal diameter, providing a magnetic field of 3.8\unit{T}. Within the solenoid volume are a silicon pixel and strip tracker, a lead tungstate crystal electromagnetic calorimeter (ECAL), and a brass and scintillator hadron calorimeter (HCAL), each composed of a barrel and two endcap sections. Forward calorimeters extend the pseudorapidity ($\eta$) coverage provided by the barrel and endcap detectors. Muons are reconstructed using gas-ionization detectors embedded in the steel flux-return yoke outside the solenoid. More detailed descriptions of the CMS detector, together with a definition of the coordinate system used and the relevant kinematic variables, can be found in Refs.~\cite{CMS:2008xjf,CMS:2023gfb}.

At the start of 2017, a new pixel detector was installed~\cite{Phase1Pixel}; the upgraded tracker measured charged particles within the range $\abs{\eta} < 3.0$, covering the radius between 4 and 110\cm. During the LHC running period when the data used in this paper were recorded, the silicon tracker consisted of 1856 silicon pixel and 15\,148 silicon strip detector modules. For nonisolated particles with transverse momentum (\pt) in the range $1 < \pt < 10\GeV$ and $\abs{\eta} < 3.0$, the track resolutions are typically 1.5\% in \pt and 20--75\mum in the transverse impact parameter (\dxy)~\cite{DP-2020-049,CMS:2014pgm}.

Events of interest are selected using a two-tiered trigger system. The first level, referred to as the level-1 (L1) trigger, is composed of custom hardware processors and uses information from the calorimeters and muon detectors to select events at a rate of around 100\unit{kHz} within a fixed latency of about 4\mus~\cite{CMS:2020cmk}. The second level, known as the high-level trigger, consists of a farm of processors running a version of the full event reconstruction software optimized for fast processing and reduces the event rate to around 1\unit{kHz} before data storage~\cite{CMS:2016ngn,CMS:2020cmk}.

A particle-flow (PF) algorithm~\cite{CMS:2017yfk} aims to reconstruct and identify each individual particle in an event, with an optimized combination of information from the various elements of the CMS detector. The energy of photons is obtained from the ECAL measurement. The energy of electrons is determined from a combination of the electron momentum at the primary interaction vertex, as determined in the tracker, the energy of the corresponding ECAL cluster, and the energy sum of all bremsstrahlung photons spatially compatible with originating from the electron track. The energy of muons is obtained from the curvature of the corresponding track. The energy of charged hadrons is determined from a combination of their momentum measured in the tracker and the matching ECAL and HCAL energy deposits, corrected for the response function of the calorimeters to hadronic showers. Finally, the energy of neutral hadrons is obtained from the corresponding corrected ECAL and HCAL energies. The primary vertex is taken to be the vertex corresponding to the hardest scattering in the event, evaluated using tracking information alone, as described in Section 9.4.1 of Ref.~\cite{CMS-TDR-15-02}.

For each event, hadronic jets are clustered from PF candidates using the infrared- and collinear-safe anti-\kt algorithm~\cite{Cacciari:2008gp, Cacciari:2011ma} with a distance parameter of 0.4. Jet momentum is determined as the vectorial sum of all particle momenta in the jet, and is found from simulation to be, on average, within 5--10\% of the true momentum over the entire \pt spectrum and detector acceptance. Additional proton-proton interactions within the same or nearby bunch crossings (referred to as pileup) can contribute additional tracks and calorimetric energy depositions to the jet momentum. To mitigate this effect, charged particles identified as originating from pileup vertices are discarded, and an offset correction is applied to correct for remaining contributions~\cite{CMS:2020ebo}. Jet energy corrections are derived from simulation to bring the measured response of jets to that of particle-level jets on average. In situ measurements of the momentum balance in dijet, photon+jet, {\PZ}+jet, and multijet events are used to account for any residual differences in the jet energy scale between data and simulation~\cite{CMS:2016lmd}. The jet energy resolution amounts typically to 15--20\% at 30\GeV, 10\% at 100\GeV, and 5\% at 1\TeV~\cite{CMS:2016lmd}. Additional selection criteria are applied to remove jets potentially dominated by anomalous contributions from various subdetector components or reconstruction failures~\cite{CMS:2017wyc}. To identify jets originating from \PQb quark fragmentation (\PQb jets), the ``tight'' working point of the \textsc{DeepJet} tagging algorithm is used, which has an identification efficiency for \PQb jets from top quark decays with $\pt > 30\GeV$ of about 58\% and a misidentification probability for light jets (from the fragmentation of \PQu, \PQd, \PQs quarks and gluons) of about 0.1\%~\cite{CMS:2017wtu, Bols:2020bkb, CMS-DP-2023-005}. 

The \ptvecmiss is computed as the negative vector \pt sum of all the PF candidates in an event, and its magnitude is denoted as \ptmiss~\cite{CMS:2019ctu}. The \ptvecmiss is modified to account for corrections to the energy scale of the reconstructed jets in the event. Anomalous events with high \ptmiss can be due to a variety of reconstruction failures, detector malfunctions, or noncollision backgrounds. Such events are rejected by event filters that are designed to identify more than 85--90\% of the spurious high-\ptmiss events with a mistagging rate less than 0.1\%~\cite{CMS:2019ctu}.

\begin{sloppypar}Hadronic \PGt lepton decays (\tauh) are reconstructed from jets using the hadrons-plus-strips algorithm~\cite{CMS:2018jrd}, which combines one or three tracks with energy deposits in the calorimeters to identify the \PGt decay modes.
\end{sloppypar}

\section{Preselection and event samples}
\label{sec:datasets}

Events used in this search are required to pass a preselection, designed to select events with a high-\pt ISR jet and large \ptmiss. These events are recorded by a trigger requiring $\ptmiss > 120\GeV$ and are required to have offline $\ptmiss > 400\GeV$, which removes background events, while retaining signal events. In addition, events are required to have at least one jet with $\pt > 100\GeV$ and $\abs{\eta}<2.4$. The highest \pt jet in an event is considered the ISR candidate jet. The preselection also rejects events containing well-reconstructed muon, electron, photon, or hadronically decaying \PGt candidates that satisfy the following requirements. Muons must have $\pt > 10\GeV$ and $\abs{\eta} < 2.4$~\cite{CMS:2018rym}. Electrons are required to have $\pt > 10\GeV$ and $\abs{\eta} < 2.5$~\cite{CMS:2020uim}. Photons must have $\pt > 15\GeV$ and $\abs{\eta} < 2.5$~\cite{CMS:2020uim}. Tau leptons are required to have $\pt > 18\GeV$ and $\abs{\eta} < 2.3$~\cite{CMS:2018jrd}. No requirements are imposed on the degree of detector activity in the vicinity of any such lepton or photon. This veto allows for the amount of background to be reduced by a factor of two. 

Signal events are simulated at leading order (LO) using \MGvATNLO~2.6.5~\cite{Alwall:2014hca}, with up to two additional jets included in the matrix element calculations. The MLM~\cite{Alwall:2007fs} prescription is used to match jets from matrix element calculations to those from parton showers. The LLP production cross sections are normalized to next-to-next-to-LO (NNLO) accuracy in quantum chromodynamics (QCD) and matched to next-to-next-to-leading logarithmic (NNLL) soft-gluon resummation~\cite{Beenakker:2016lwe}. For the \PSQt NLSP model, bound states resulting from the color charges carried by \PSQt could have kinematic effects and interact with materials in the detector. The interactions follow the cloud model~\cite{Kraan:2004tz, Mackeprang:2006gx} in the simulation.
The \PSQt NLSP samples are generated with different \PSQt masses ranging from 400 to 1400\GeV, with \dm ranging from 12 to 25\GeV, and \br ranging from 10 to 100\%, corresponding to \ctau in the range of 0.01--400\mm. The lower bound on \dm is set to 12\GeV, as smaller values of \dm correspond to \ctau ranges where the search sensitivity diminishes. The bino-wino NLSP samples are generated with different masses of the NLSP ranging from 200 to 600\GeV, with \dm ranging from 12 to 25\GeV, and \ctau ranging from 0.2 to 200\mm.

As a result of the significant \ptmiss requirement, the dominating background processes include \PZ and \PW boson production in association with up to four jets from the matrix element calculations ($\PZ/\PW$+jets), where the \PZ boson decays to neutrino-antineutrino pairs and the \PW boson decays to a lepton and its corresponding neutrino. Additional contributions arise from \ttbar, events with jets produced through the strong interaction (QCD events), and single top quark production. The $\PZ/\PW$+jets and QCD events are generated using \MGvATNLO at LO, with the MLM~\cite{Alwall:2007fs} prescription for jet matching applied. The \ttbar events are generated using \MGvATNLO 2.6.1 at next-to-LO (NLO), with the FxFx~\cite{Frederix:2012ps} prescription for jet matching. Single top quark production is generated using \POWHEG~\cite{Nason:2004rx, Frixione:2007vw, Alioli:2009je, Alioli:2010xd, Re:2010bp} (version 2.0 for $t$-channel and 1.0 for $\PQt\PW$-channel production) at NLO in QCD, and the diboson processes are generated with \PYTHIA 8.240~\cite{Sjostrand:2014zea}. Simulated background events are used solely for selection optimization and systematic uncertainty studies and are not employed in the search results, as the background is evaluated from data.

The NNPDF3.1 NNLO~\cite{NNPDF:2014otw,NNPDF:2017mvq} set of parton distribution functions (PDFs) is used for all simulated samples. The \MGvATNLO and \POWHEG generators are interfaced with \PYTHIA with the underlying event tune CP5~\cite{CMS:2019csb} to model parton showering and fragmentation.

For all the simulated events, \GEANTfour~\cite{GEANT4:2002zbu} is used to simulate the CMS detector response. To describe the effect of pileup, simulated minimum bias events are superimposed on the hard interactions, with the multiplicity matching that observed in data.

\section{Displaced vertex reconstruction}
\label{sec:vtxreco}

Displaced vertices are found using well-reconstructed and displaced tracks. A track is considered well reconstructed if it has $\pt> 0.5\GeV$, at least six associated tracker hits, a normalized $\chi^2 < 5$, which represents the compatibility between the measured tracker hit positions and the positions predicted by the fitted track, and a relative uncertainty in \pt less than 1.5\%. The displacement criterion requires a track to satisfy \dxysig to be at least 2, where \dxysig is defined as the ratio of the \dxy to its associated uncertainty and is calculated with respect to the primary vertex.

Displaced vertices are reconstructed by the inclusive vertex finder (IVF)~\cite{CMS:2017wtu}, which is based on the adaptive vertex fitter~\cite{DHondt:2004ipa}. In the IVF method, the selected tracks are sorted into multiple clusters based on the distance of the closest approach between tracks. The adaptive vertex fitter is then applied iteratively to each cluster to find the candidates for displaced vertices. During the fitting process, multiple displaced vertices can be fit in a single cluster, and tracks are assigned to vertices with weights reflecting the likelihood that each track originates from a given vertex. A deterministic annealing algorithm is applied to obtain a robust fit result. Vertex candidates are further merged if they share more than 70\% of their associated tracks and if their separation satisfies $d_{\text{vv}}/\sigma_{d_{\text{vv}}}<2$, where $d_{\text{vv}}$ is the distance between the vertices and $\sigma_{d_{\text{vv}}}$ is its associated uncertainty. To obtain optimally reconstructed vertices, an arbitration step is performed to assign tracks to vertex candidates when they are in close proximity. In this step, the distance between unassociated tracks and reconstructed vertices is evaluated. Tracks found to be sufficiently close to a vertex are added to it, followed by a vertex refit. This procedure effectively ``recycles'' tracks that were not already included in vertices, thereby enhancing the overall vertex reconstruction performance. Following the track arbitration, an additional merging step is performed to prevent track sharing: vertex candidates are merged if their distance divided by their uncertainty is less than 10 and they share more than 20\% of their tracks. Each time the set of tracks in a vertex is updated, the vertex is refitted.

\begin{figure}[htbp]
    \centering
    \includegraphics[width=0.8\textwidth]{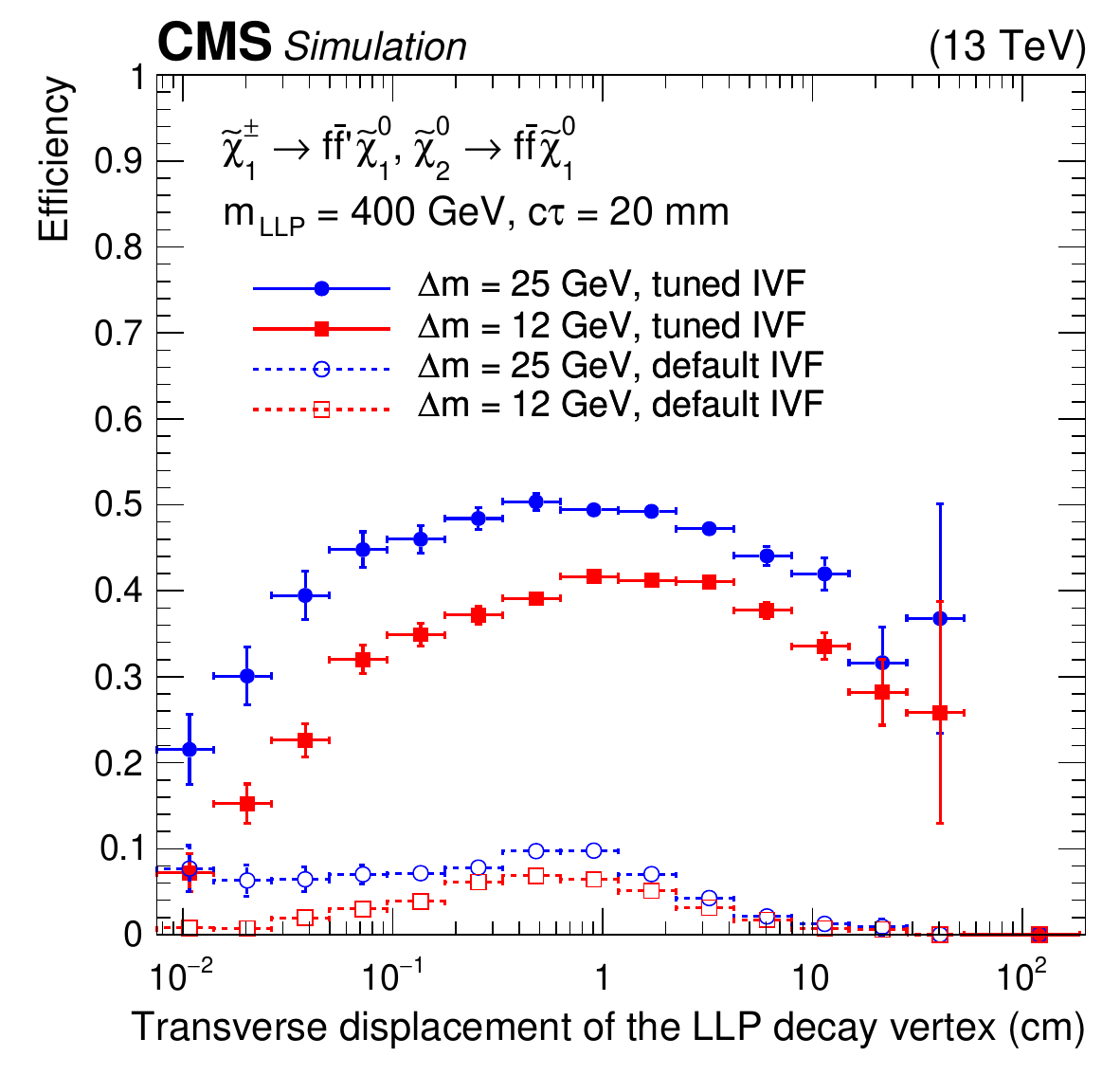}
    \caption{The LLP reconstruction efficiency as a function of the transverse displacement of the LLP decay vertex. Bino-wino NLSP samples with $\mllp = 400\GeV$, $\ctau = 20\mm$, and $\dm = 25$ (blue circles) and 12 (red squares)\GeV are shown in the plot. Tuned (solid) and default (dashed) IVF are compared.}
    \label{fig:llpeff}
\end{figure}

For this search, parameters of the IVF are tuned to allow vertices with tracks that have large angular separation, thus enhancing the reconstruction efficiency for LLP decays with large opening angles between the decay products and large \ptmiss. The reconstruction efficiency, defined as the fraction of LLPs that are reconstructed as displaced vertices, is shown in Fig.~\ref{fig:llpeff} as a function of LLP transverse displacement. The results are presented for bino-wino NLSP samples with an LLP mass (\mllp) of 400\GeV, $\ctau = 20\mm$, and $\dm = 25$ and 12\GeV. The tuned IVF shows significantly improved efficiency compared with the default configuration across both \dm values and displacement ranges. The reconstruction efficiency rises with increasing transverse displacement up to a few centimeters, then gradually decreases because of reduced tracking and vertexing performance at larger displacements.

\begin{figure}[htbp]
    \centering
    \includegraphics[width=0.8\textwidth]{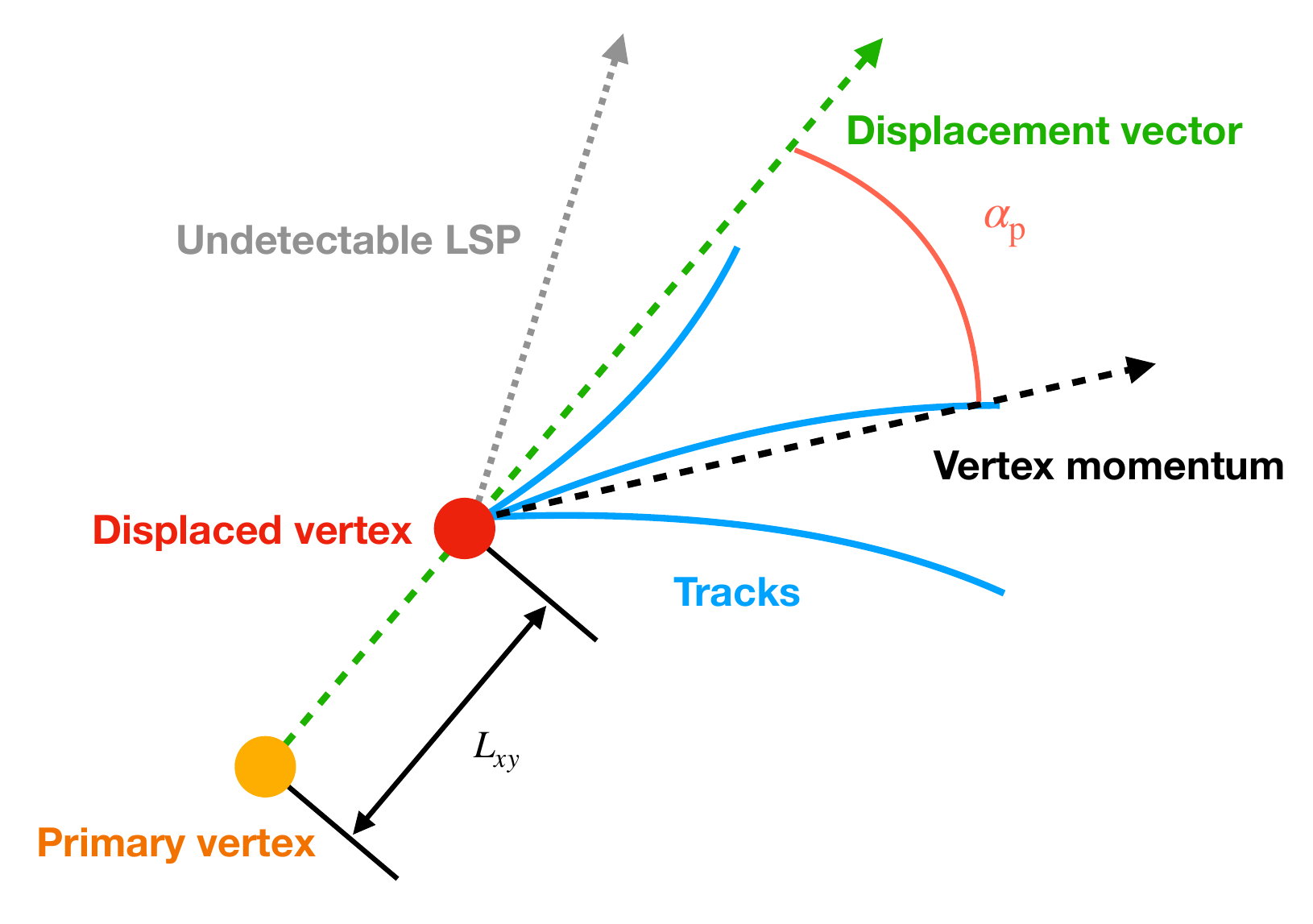}
    \caption{Diagram that shows the key features of a displaced vertex.}
    \label{fig:dv}
\end{figure}

Figure~\ref{fig:dv} shows the key features of a displaced vertex. The transverse distance between the primary and displaced vertices is referred to as \lxy. The displacement vector is defined as the vector pointing from the primary to the displaced vertex. The vertex momentum is calculated as the vector sum of the momenta of all tracks at the vertex. The three-dimensional angle between the displacement vector and the vertex momentum is referred to as \alphap. 

The reconstructed displaced vertices can include background vertices resulting from the SM LLPs, such as \PQb hadrons, nuclear interactions between particles and materials in the detector, and random crossings of unrelated tracks.  Background vertices are removed by applying selection criteria to both the tracks associated with the vertices and the vertices themselves.

A ``good track'' must satisfy the following criteria. The number of associated detector hits measured in the tracker must be greater than 13, which ensures that the track is well measured. To remove tracks from pileup vertices, the absolute value of the ratio between \dxy and the longitudinal impact parameter (\dz) must be greater than 0.25. Given the recoil of the ISR jet, LLP decay products tend to be in the same hemisphere as the \ptvecmiss and the opposite hemisphere of the ISR jet. As a result, the difference in the azimuthal angle ($\phi$) between the track and \ptvecmiss ($\Delta \phi(\text{trk,\,\ptvecmiss})$) is required to be less than 1.5 radians, while the $\phi$ difference between the track and the ISR jet candidate ($\Delta \phi(\text{trk,\,ISR})$) is required to be greater than 1. To reject tracks within jets, we calculate the track isolation, defined as the scalar sum of the transverse energy of charged hadrons, neutral hadrons, and photons within a cone of $\DR= 0.3$, where $\DR \equiv \sqrt{\smash[b]{(\dphi)^2+(\deta)^2}}$ corresponds to the angular distance between the track and each particle considered in the sum. The ratio of the track isolation to the track \pt is required to be less than 5. In addition, we require tracks to have \dxysig greater than 4 to further remove prompt background tracks.

\begin{figure}[htbp]
    \centering
    \includegraphics[width=0.8\textwidth]{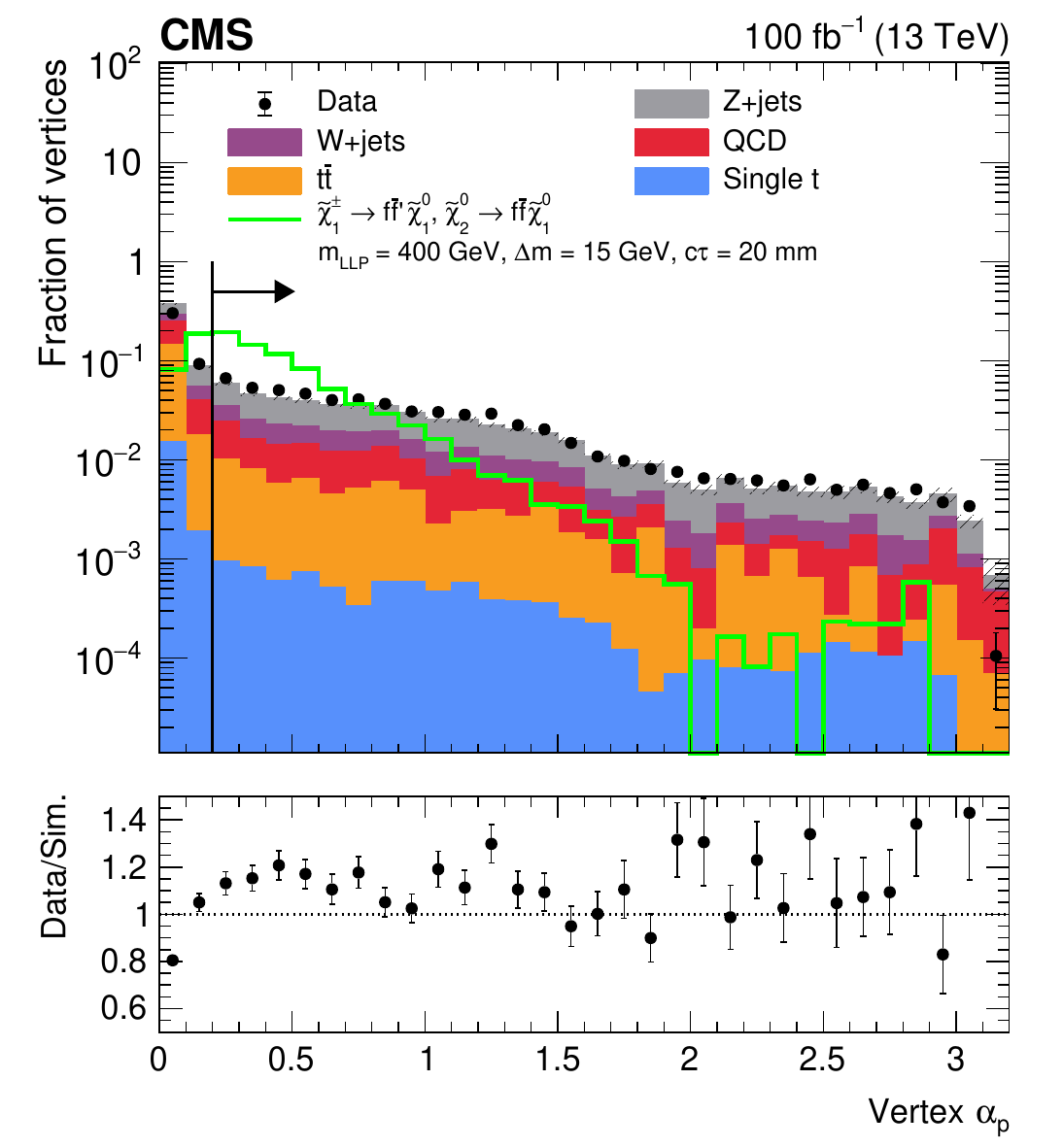}
\caption{The vertex \alphap distribution compared between data, simulation, and bino-wino NLSP sample with $\mllp = 400\GeV$, $\ctau = 20\mm$, and $\dm = 15\GeV$. All distributions are normalized to unity. The ratio of data to simulation is shown in the lower panel. The arrow indicates the value of the optimal \alphap threshold.}
    \label{fig:vtx_pangle}
\end{figure}

The following selections on vertex features are also applied. The number of degrees of freedom, calculated as two times the sum of track weights, less 3, must be greater than 1, which ensures the vertices are well reconstructed. Following the same argument as the good track selection, the $\phi$ difference between the vertex displacement vector and \ptvecmiss ($\Delta \phi(\text{vtx,\,\ptvecmiss})$) is required to be less than 1.5, and the $\phi$ difference between the vertex displacement vector and the ISR jet candidate ($\Delta \phi(\text{vtx,\,ISR})$) is required to be greater than 1. In addition, the number of tracks that satisfy $\abs{\dxy/\dz}$ greater than 0.25 must be at least one to remove vertices from pileup. To make the search sensitive to LLPs with different lifetimes, no explicit selection on vertex \lxy is applied. Figure~\ref{fig:vtx_pangle} compares the \alphap distribution for vertices that satisfy the aforementioned selections in data, simulated background, and signal from the bino-wino NLSP sample with $\mllp = 400\GeV$, $\ctau = 20\mm$, and $\dm = 15\GeV$. It shows that a large fraction of background vertices have a value of \alphap close to 0. A selection of $\alphap > 0.2$ is imposed. This threshold is optimized to maximize the search sensitivity based on the event yields in the SRs and is not directly derived from Fig.~\ref{fig:vtx_pangle}. To reduce model dependence, no upper bound is imposed on \alphap. Comparing data and background simulation, we find that data vertices tend to have larger \alphap. This discrepancy arises from the mismodeling of pileup, as vertices in data are more likely to include additional tracks originating from pileup interactions, which in turn lead to higher \alphap values. 

\begin{figure}[htbp]
    \centering
    \includegraphics[width=0.49\textwidth]{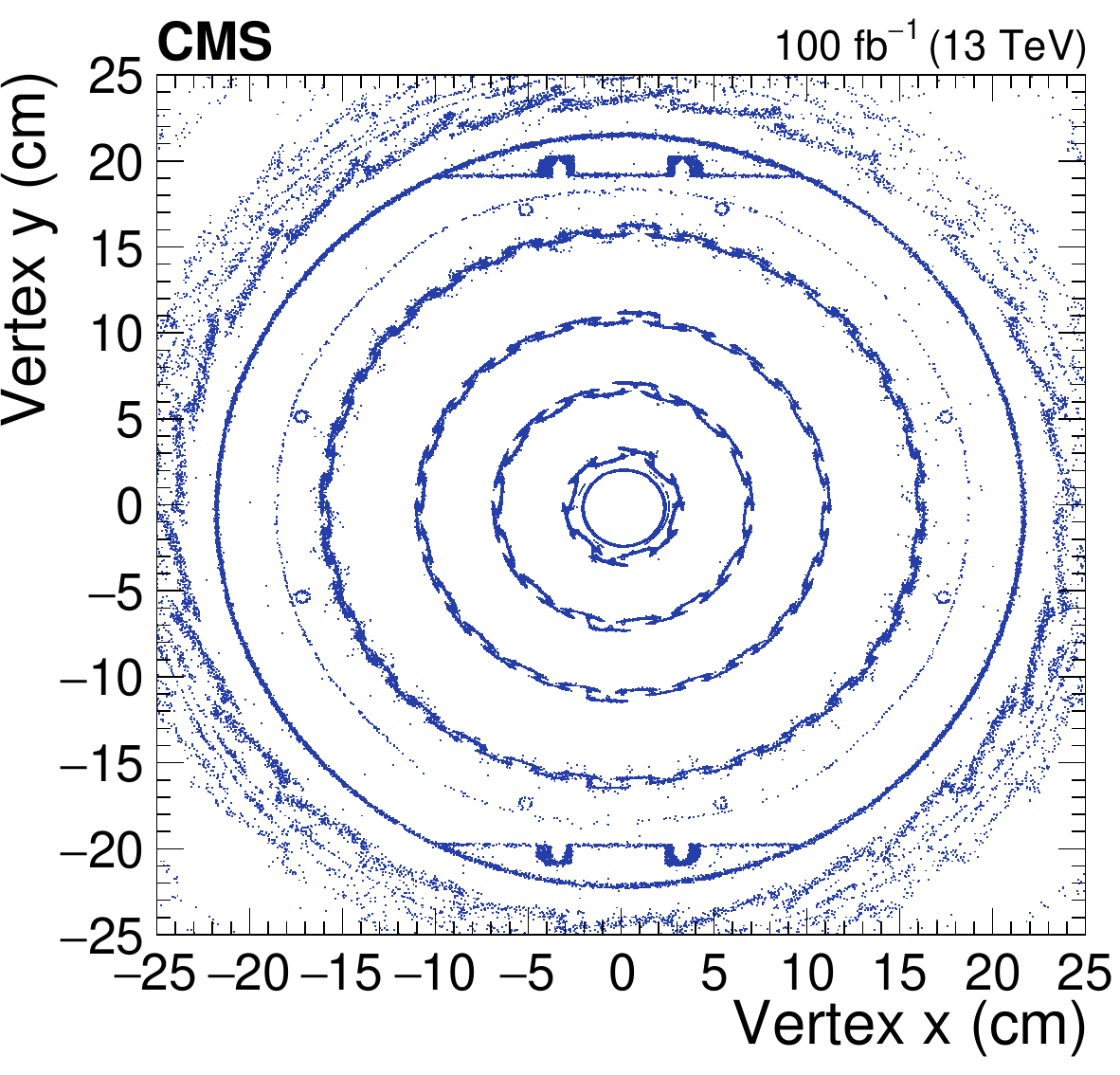}
    \caption{Material map for the CMS tracker derived from data. A zoomed-in view is provided for $x$ and $y$ within the range $-25$ to $25$\cm.}
    \label{fig:map}
\end{figure}

To remove vertices that result from nuclear interactions of SM particles with the tracker material, we veto vertices whose positions are consistent with originating within the tracker material. To this end, we derived a material map from the positions of displaced vertex candidates in 2017 and 2018 data. Events that satisfy the \ptmiss trigger with a minimum threshold of 120\GeV and pass the noise filters~\cite{CMS:2019ctu} are used for this purpose. To avoid potential bias, events that contribute to the SRs defined in Section~\ref{sec:bkgest} are excluded in the derivation of the map. In each event, displaced vertices are reconstructed and required to have at least two tracks and greater than one degree of freedom. The map is obtained by projecting all of the selected vertices onto the plane transverse to the beam. To remove vertices from SM LLP decays and random track crossings, a requirement on the vertex density as a function of the transverse radius is applied to exclude regions with fewer vertices. The resulting map is shown in Fig.~\ref{fig:map}. A similar study was performed in Ref.~\cite{CMS:2020iwv}, and the resulting maps are found to be consistent. The map is applied to both data and simulated events. The material map veto reduces the signal efficiency by 0--10\%, depending on the lifetime of the LLP.

Vertices that meet the criteria based on the vertex features and are not vetoed by the material map are said to have passed the vertex selections and are used in the analysis. 

\section{Event selection and background estimation}
\label{sec:bkgest}

\begin{figure}[htbp]
    \centering
    \includegraphics[width=0.8\textwidth]{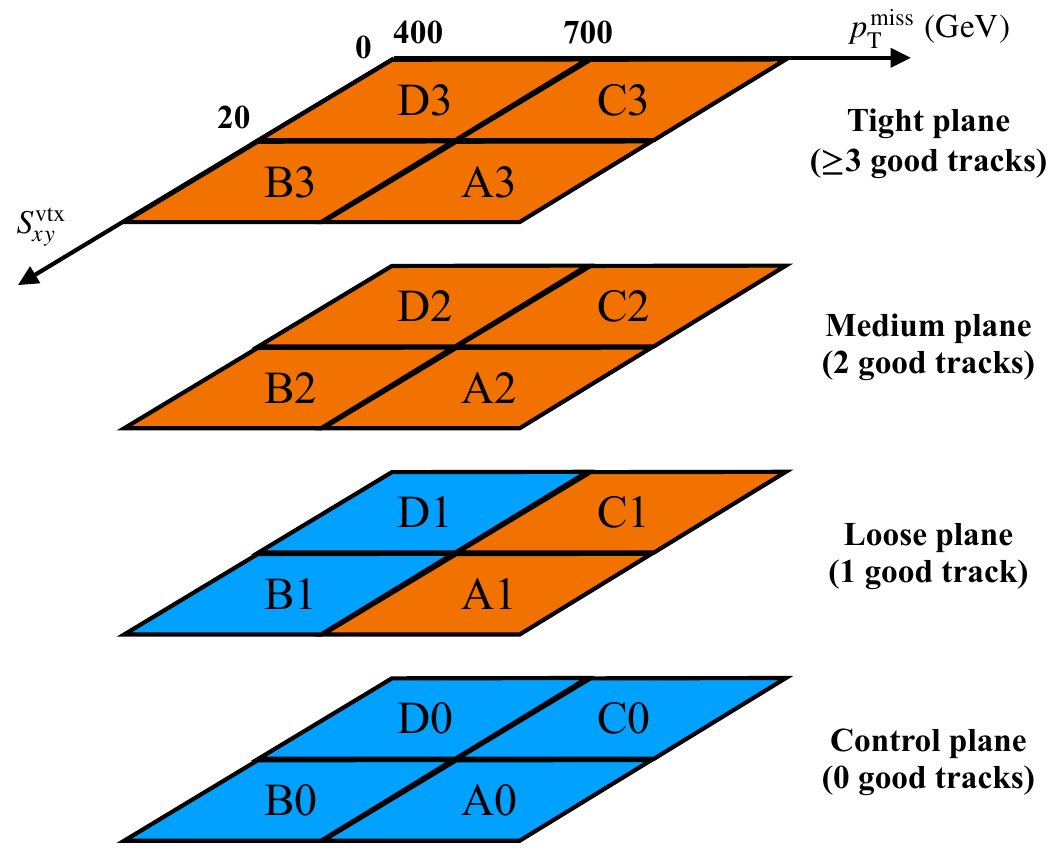}
    \caption{Definition of the signal (orange) and control (blue) regions. Different planes are defined based on the number of good tracks in the vertex. In each plane, different regions are divided according to the \ptmiss and \lxysig values. The letters in the boxes correspond to the region labels described in the text, while the numbers in the boxes correspond to the plane numbers.}
    \label{fig:regions}
\end{figure}

In addition to the preselection defined in Section~\ref{sec:datasets}, events with at least one vertex that satisfies the vertex selections are further divided into four exclusive categories based on the number of good tracks. Events containing at least one vertex with at least three good tracks are assigned to the tight signal plane. Those with at least one vertex containing exactly two good tracks, and not satisfying the tight signal plane criteria, are assigned to the medium signal plane. The remaining events with at least one vertex with one good track are assigned to the loose signal plane. Events with vertices with no good tracks are classified into the control plane. Within each plane, four regions are defined based on whether the requirements $\ptmiss > 700\GeV$ and $\lxysig > 20$, where \lxysig denotes the ratio of the vertex transverse displacement \lxy to its uncertainty, are satisfied. The \lxysig and \ptmiss variables are found to be statistically independent for the background processes. The selections are optimized to achieve the maximum sensitivity to the benchmark signal models. In events with more than one vertex, the vertex with the greatest \lxysig is used. The definition of the regions is illustrated in Fig.~\ref{fig:regions}. The SRs include all regions in the tight and medium signal plane, and regions A and C in the loose signal plane. All other regions are used as control regions (CRs), including regions B and D in the loose signal plane and all regions in the control plane. 

This search adopts a novel background estimation method using transfer factors, which is purely based on data. The overall strategy consists of multiplying the number of events in the control plane by transfer factors to predict the number of events in other planes. Because of the $\dxysig>4$ requirement in the good track selection, \lxysig and the number of good tracks are statistically dependent. To mitigate the impact of this dependence, the background estimation is performed without the \lxysig selection. Instead, predictions are made for a given \ptmiss range and the number of good tracks to ensure a more reliable estimation.

\begin{figure}[htbp]
    \centering
    \includegraphics[width=0.8\textwidth]{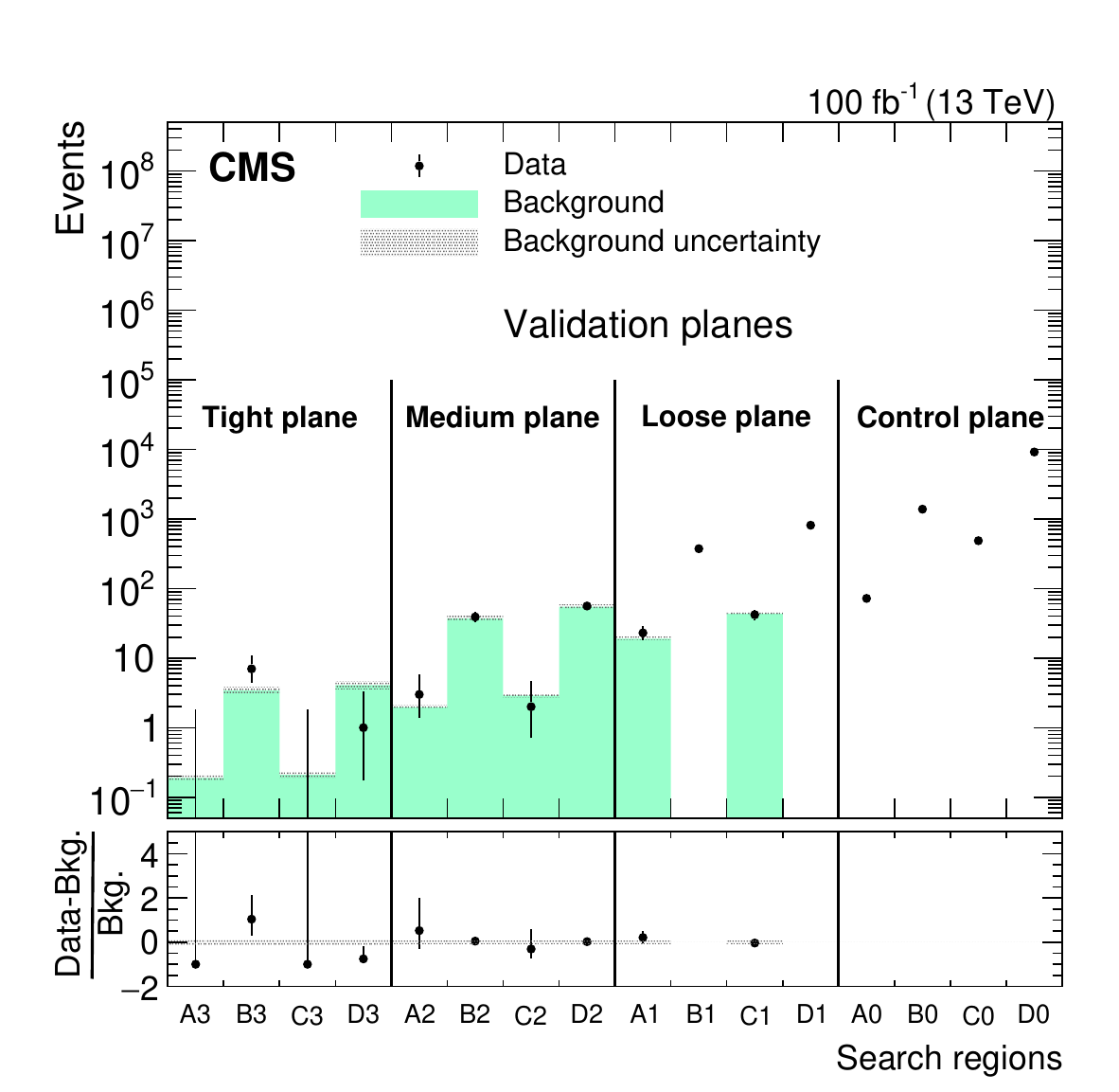}
    \caption{The number of observed and predicted background events in the regions of the validation planes. The predicted background is shown with its associated uncertainties. The observed data are displayed with the 68\% Poisson intervals. The lower panel shows the fractional difference between the observed data and the predicted background.}
    \label{fig:validation}
\end{figure}

To predict the number of events without \lxysig selection, a transfer factor ($f_{i\to j}$) is defined as the ratio of events in the target plane with $j$ good tracks to those in the original plane with $i$ good tracks. These transfer factors are computed using events in the low-\ptmiss region ($200 < \ptmiss < 400 \GeV$), as shown by Eq.~\eqref{eq:1}. The transfer factor for transitioning from zero to one good track is derived from CRs B0, D0, B1, and D1 in the nominal search planes, as shown by Eq.~\eqref{eq:2}. Additionally, the transfer factors obtained from low-\ptmiss events are scaled by the ratio of transfer factors between the search planes (zero to one good track) and the low-\ptmiss planes, as shown by Eq.~\eqref{eq:3}. The estimated event yield is then obtained by multiplying the number of events in the CRs by the corresponding transfer factors, as shown by Eqs.~\eqref{eq:4} and~\eqref{eq:5}.
\begin{align}
    f_{i\to j}^{\text{low-\ptmiss}} &= \frac{N_{\text{bkg}}^{j\,\text{good track}}}{N_{\text{bkg}}^{i\,\text{good track}}}&&(i,j)=(0,1),(1,2),(2,3).\label{eq:1} \\
    f_{0\to 1} &= \frac{N_{\text{bkg}}^{\text{B1}+\text{D1}}}{N_{\text{bkg}}^{\text{B0}+\text{D0}}}. \label{eq:2} \\
    f_{i\to i+1} &= \frac{f_{0\to1}}{f_{0\to1}^{\text{low-\ptmiss}}}f_{i\to i+1}^{\text{low-\ptmiss}}\,&&i=1,2.\label{eq:3} \\
    N_{\text{bkg}}^{\text{A}i+\text{C}i} &= N_{\text{bkg}}^{\text{A0}+\text{C0}} \prod_{j=0}^{i-1} f_{j\to j+1}\,&&i=1,2,3.\label{eq:4} \\
    N_{\text{bkg}}^{\text{B}i+\text{D}i} &= N_{\text{bkg}}^{\text{B1}+\text{D1}} \prod_{j=1}^{i-1} f_{j\to j+1}\,&&i=2,3.\label{eq:5}
\end{align}
After obtaining event counts for a given \ptmiss range and number of good tracks, the prediction for events with the \lxysig selection is made. The \lxysig distributions for each plane are extracted from low-\ptmiss events ($200 < \ptmiss < 400 \GeV$) in data and used as proxies for the nominal search planes. Studies indicate that the \lxysig distribution for a given plane remains consistent across different \ptmiss ranges, justifying the use of low-\ptmiss events as a representative sample. From these distributions, for a given plane with $i$ good tracks, the fractions of events satisfying the \lxysig selection ($f_{i}^{\text{vtx}}$) are computed. The final background prediction is obtained by multiplying this fraction by the previously estimated total number of events, yielding predictions across all SRs. The calculations are shown by Eqs.~(\ref{eq:6}--\ref{eq:9}).
\begin{align}
    N_{\text{bkg}}^{\text{A}i} &= N_{\text{bkg}}^{\text{A}i+\text{C}i}f_{i}^{\text{vtx}}\,&&i=1,2,3.\label{eq:6} \\
    N_{\text{bkg}}^{\text{C}i} &= N_{\text{bkg}}^{\text{A}i+\text{C}i}\left(1-f_{i}^{\text{vtx}}\right)\,&&i=1,2,3. \label{eq:7} \\
    N_{\text{bkg}}^{\text{B}i} &= N_{\text{bkg}}^{\text{B}i+\text{D}i}f_{i}^{\text{vtx}}\,&&i=2,3. \label{eq:8} \\
    N_{\text{bkg}}^{\text{D}i} &= N_{\text{bkg}}^{\text{B}i+\text{D}i}\left(1-f_{i}^{\text{vtx}}\right)\,&&i=2,3. \label{eq:9}
\end{align}
To validate the background estimation method, orthogonal validation planes are constructed by applying different requirements of $\Delta \phi(\text{trk/vtx,\,\ptvecmiss/ISR})$. Specifically, it is required that $\Delta \phi(\text{trk/vtx,\,\ptvecmiss}) > 1.5$ and $\Delta \phi(\text{trk/vtx,\,ISR})>0.4$. These selections ensure that there is no overlap with the SR and that the vertex is not within the ISR jet area. Figure~\ref{fig:validation} compares the observation and prediction for regions in the validation planes. It presents the predicted central values along with their associated uncertainties. In the CRs, no predictions are made. For most of the regions in the validation planes, the prediction and observation agree within statistical uncertainties.

\section{Systematic uncertainties}
\label{sec:syst}

The impact of systematic uncertainties associated with the background estimation and the signal efficiency, expressed as the resulting changes in the event yields, is summarized in Table~\ref{tab:signal_syst_unc}.

As described in Section~\ref{sec:bkgest}, the background is estimated from data. Therefore, the modeling of the background in simulation has no impact on the analysis. For $f_{0\to1}$ and $f_{i}^{\text{vtx}}$, the statistical uncertainty in the derivation is used as a systematic uncertainty of up to 2\%. The prediction and observation when performing the background estimate procedure in the validation plane defined in Section~\ref{sec:bkgest} are consistent within the statistical uncertainties, so we apply no additional systematic uncertainty associated with the background estimation.

Systematic uncertainties associated with the signal efficiency result from the modeling of displaced tracks and vertex reconstruction, the accuracy of the material map, jet energy scale, jet energy resolution, unclustered energy, pileup, trigger, lepton and photon veto, integrated luminosity, and the running conditions. 

\begin{figure}[htbp]
    \centering
    \includegraphics[width=0.8\textwidth]{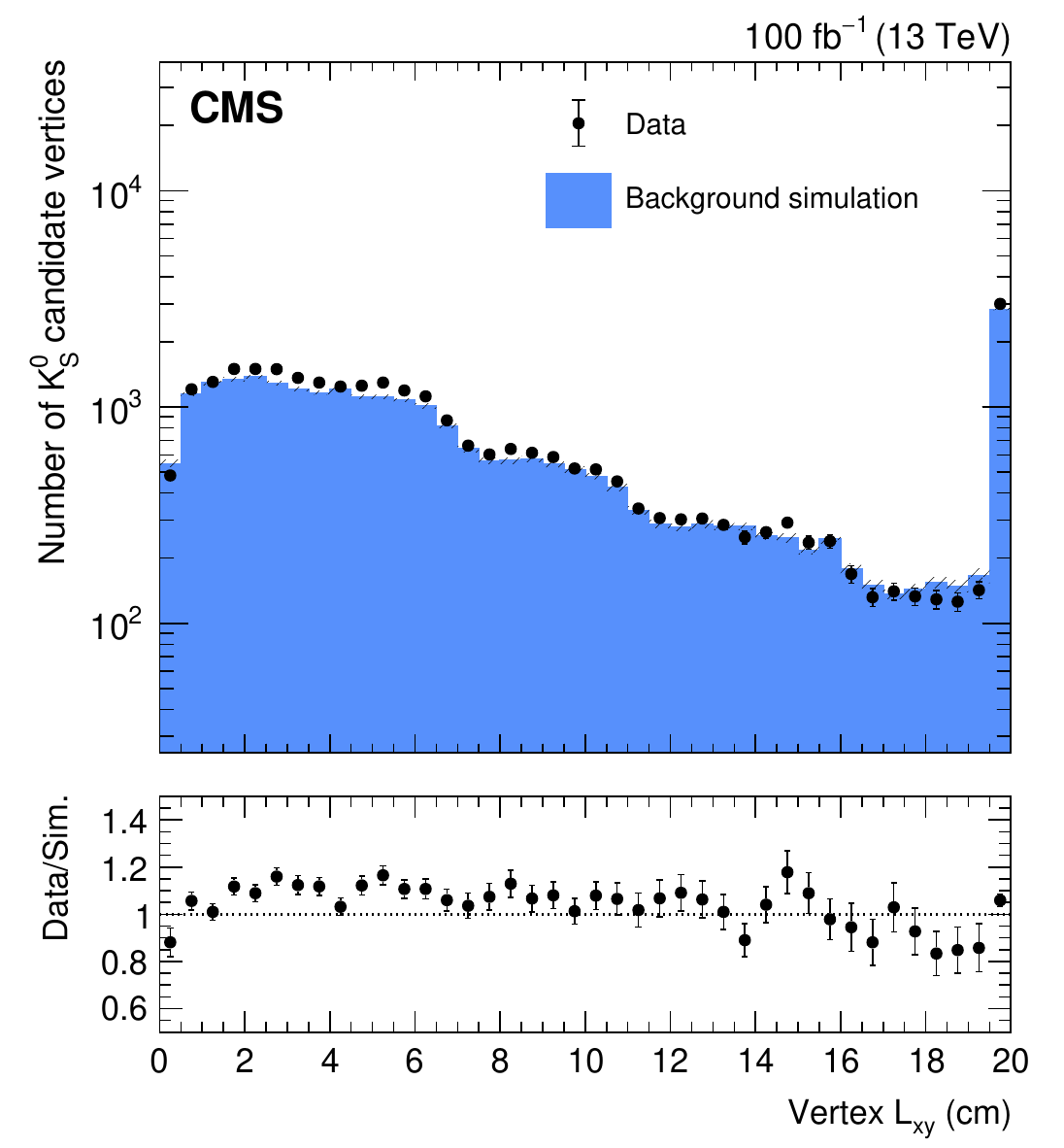}
    \caption{The \PKzS decay candidate vertex \lxy distribution compared between data and simulation. The ratios between data and simulation are shown in the lower panel. }
    \label{fig:tkmaxdxy}
\end{figure}

The displaced track and vertex reconstruction efficiency is studied by reconstructing displaced vertices from $\PKzS\to\PGpm\PGpp$ decays~\cite{CMS-DP-2024-010}. Events that satisfy the event preselection described in Section~\ref{sec:datasets} are used in the study. Vertices are reconstructed using the algorithms described in Section~\ref{sec:vtxreco}. To select \PKzS decay candidates, vertices are required to have exactly two tracks with opposite charges. In addition, the vertex \alphap is required to be less than 0.2, so the vertices used in this study do not overlap with the SRs. The distributions of vertex \lxy are compared between data and simulation. The \lxy distribution is obtained by a fit-based background subtraction performed in each \lxy bin. For each bin of \lxy, the invariant mass of a vertex, defined as the invariant mass of the two tracks assuming the charged pion mass hypothesis, is required to be between 0.440 and 0.560\GeV. Vertices with invariant masses outside the range 0.470--0.530\GeV are used as sidebands. A first order polynomial is fitted to the sideband regions, and the number of \PKzS decay candidate vertices is determined by subtracting the fitted background from the total number of vertices with invariant mass in the 0.470--0.560\GeV range. The simulation is normalized such that the number of \PKzS decay candidate vertices at small \lxy is the same in data and simulation. The distributions after normalization in 2017 and 2018 data are shown in Fig.~\ref{fig:tkmaxdxy}. The spread of the ratio of data and simulation, defined as 68\% of the size of its envelope, is assigned as the systematic uncertainty, with a magnitude of 10--11\%. To confirm that these results are valid for vertices with more than two tracks, a similar study is performed using vertices around \PQb jets, and consistent behavior is observed.

\begin{table}[htbp!]
\centering
\topcaption{Summary of the systematic uncertainties. The magnitude represents the change in the event yields.}
\begin{tabular}{l r}
Systematic uncertainty & Magnitude \\[1pt]
\hline
Background estimate & $<$2\% \\
Track/vertex reconstruction & 10--11\% \\
Material map veto & 1--3\% \\
Jet energy scale  & 2--4\% \\
Jet energy resolution  & 1--2\%  \\
Unclustered energy & 1--2\% \\
Pileup  & 3--4\% \\
Trigger & 3\% \\
Scale variations & 1--5\% \\
PDF variations & $<$1\% \\
Integrated luminosity & 2--3\%  \\
L1 trigger inefficiency & $<$1\% \\
\end{tabular}
\label{tab:signal_syst_unc}
\end{table}

Since the material maps are derived from data, discrepancies may arise because of potential differences in the modeling of detector material between data and simulation. To account for this, a corresponding material map is constructed using simulated events, and the resulting variation in the signal yield is taken as a systematic uncertainty. This uncertainty ranges 1--3\%.

The selections on the ISR jet and \ptmiss are affected by the jet energy scale, jet energy resolution, and unclustered energy. The corresponding systematic uncertainties are evaluated by shifting the parameters that describe the mismodeling by one standard deviation. The resulting magnitude is 2--4\% for the jet energy scale, 1--2\% for the jet energy resolution, and 1--2\% for the unclustered energy. The uncertainty in the modeling of the pileup distribution arises from the uncertainty in the total inelastic cross section. It is evaluated from the variation in the reweighted signal event yield when the cross section is varied within its uncertainty~\cite{CMS:2018mlc}, resulting in a magnitude of 3--4\%. The modeling of \ptmiss trigger efficiency results in a systematic uncertainty of 3\%. 

The measurement of the integrated luminosity for the 2017 and 2018 data-taking years results in an uncertainty of 2--3\%~\cite{CMS-PAS-LUM-17-004,CMS-PAS-LUM-18-002}. Variations in the PDFs and the renormalization and factorization scales used in signal simulation lead to changes in the selection efficiencies of the signal models. These uncertainties are estimated by reweighting the simulated signal events according to the variations among different NNPDF replicas~\cite{Butterworth:2015oua} and the scale choices. The resulting uncertainties in the selection efficiency are less than 1\% from PDFs and 1--5\% from the scale variations (up and down by a factor of two, with the two extreme variations excluded).
During the 2017 data-taking year, a gradual timing shift in the ECAL was not correctly accounted for in the L1 trigger, resulting in an efficiency drop for events with significant ECAL energy distributed in the high-$\eta$ region. This effect is simulated and results in a residual uncertainty in the signal yield of less than 1\%.

\section{Results and statistical interpretation}
\label{sec:results}

A maximum likelihood fit is performed in all of the regions described in Section~\ref{sec:bkgest}, under the background-only assumption, using the CMS statistical analysis tool \textsc{Combine}~\cite{CMS:2024onh}. The systematic uncertainties are treated as uncorrelated nuisance parameters in the fit. 

\begin{figure}[htbp]
    \centering
    \includegraphics[width=0.8\textwidth]{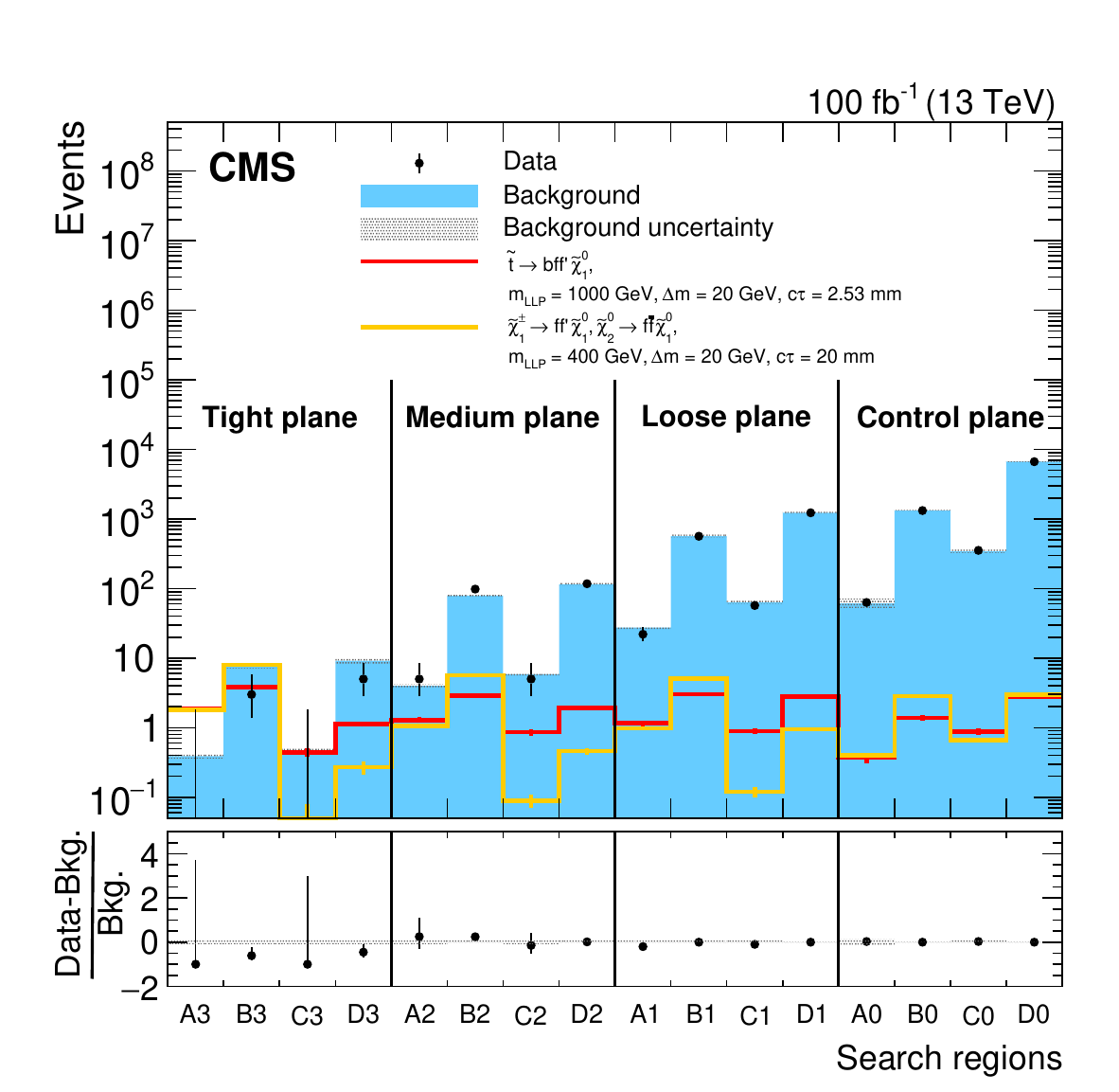}
    \caption{The number of observed and predicted background events after the fit to the regions of the search planes. In addition, two representative signals are shown. The predicted background is shown with its associated uncertainties. The observed data are displayed with the 68\% confidence level Poisson confidence intervals. The lower panel shows the fractional difference between the observed data and the predicted background.}
    \label{fig:event_yield}
\end{figure}

For each search plane, the observed data yields and background predictions after the fit are presented in Fig.~\ref{fig:event_yield}. The figure also includes the expected yields and their statistical uncertainties from two representative signal models: a \PSQt NLSP signal model with $m_{\PSQt} = 1000\GeV$, $\dm = 20\GeV$, and $\mathcal{B} = 100\%$ (\ctau=2.53\mm); and a bino-wino NLSP model with $\mllp = 400\GeV$, $\dm = 20\GeV$, and $\ctau = 20\mm$. The predicted central values are shown along with their associated uncertainties, excluding Poisson statistical uncertainties. 
Differences between the predicted and observed yields are observed in several regions. In the tight selection plane, regions B and D show predicted yields of 7.7 and 9.2 events, whereas the corresponding observed numbers of events are three and five. In region B of the medium plane, the observed yield of 98 events exceeds the prediction of 78.8. An observed $p$-value~\cite{Cousins:2018tiz} of 0.5, obtained from the combined comparison of observed and predicted background yields across all regions, indicates no significant deviation from the background-only hypothesis. Regions A and B in the tight selection plane provide the highest sensitivity for signal models with $\ctau > 1\mm$, while regions C and D are more sensitive to models with smaller \ctau values. 

To constrain signal model parameters, the 95\% confidence level (\CL) upper limit on the production cross section for a given assumption of the branching fraction is extracted using the \CLs criterion~\cite{CLS2,CLS1,CMS:2024onh} with the profile likelihood ratio as the test statistic. Data sets from different data-taking years are treated separately and combined during the final fit. The upper limits are compared with the theoretical prediction of the production cross section and uncertainty~\cite{Borschensky:2014cia} calculated at NNLO+NNLL precision~\cite{Beenakker:2016lwe} to exclude regions of the parameter space of the benchmark signal models. 

\begin{figure}[htbp]
    \centering
    \includegraphics[width=0.45\textwidth]{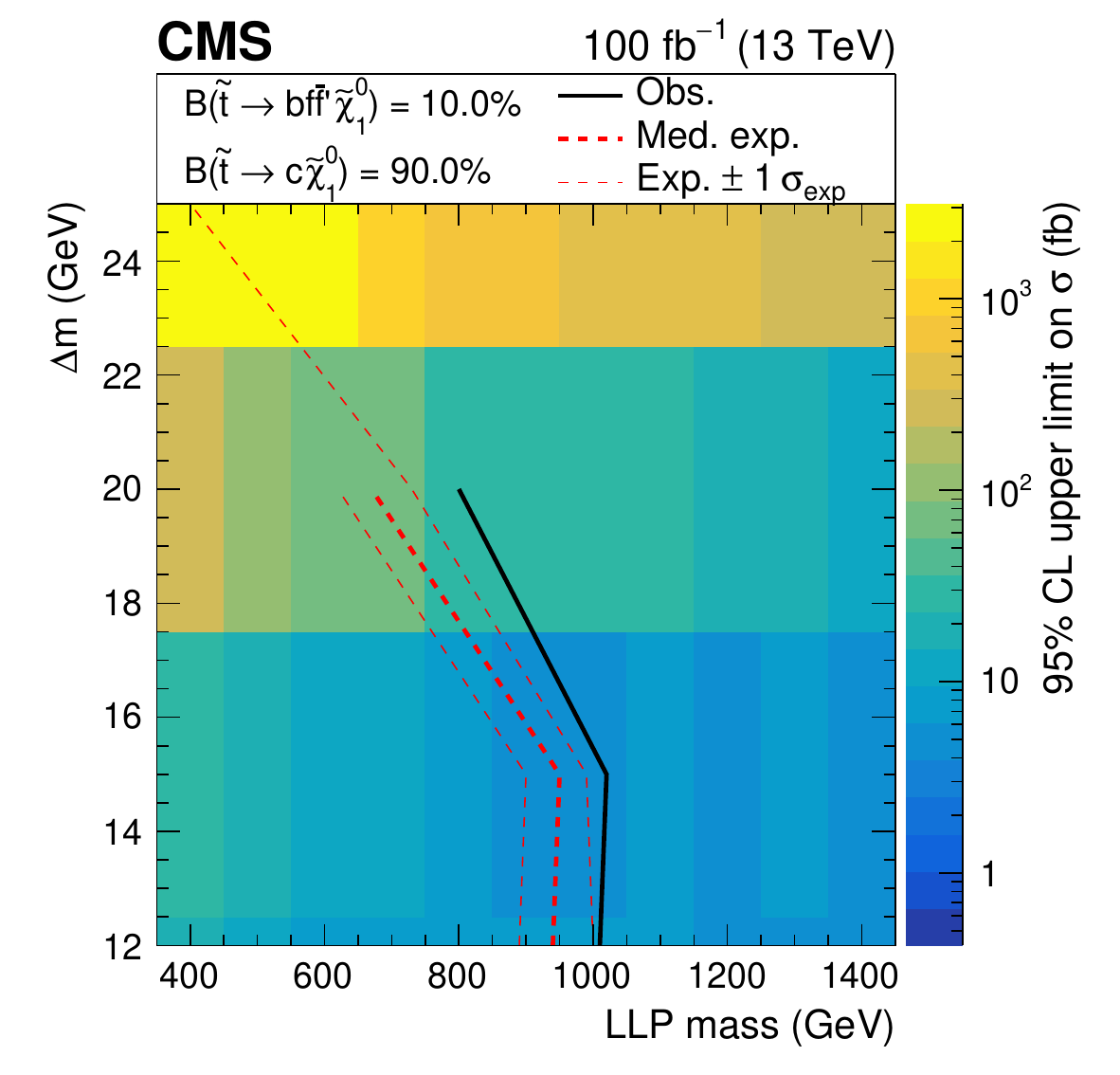}
    \includegraphics[width=0.45\textwidth]{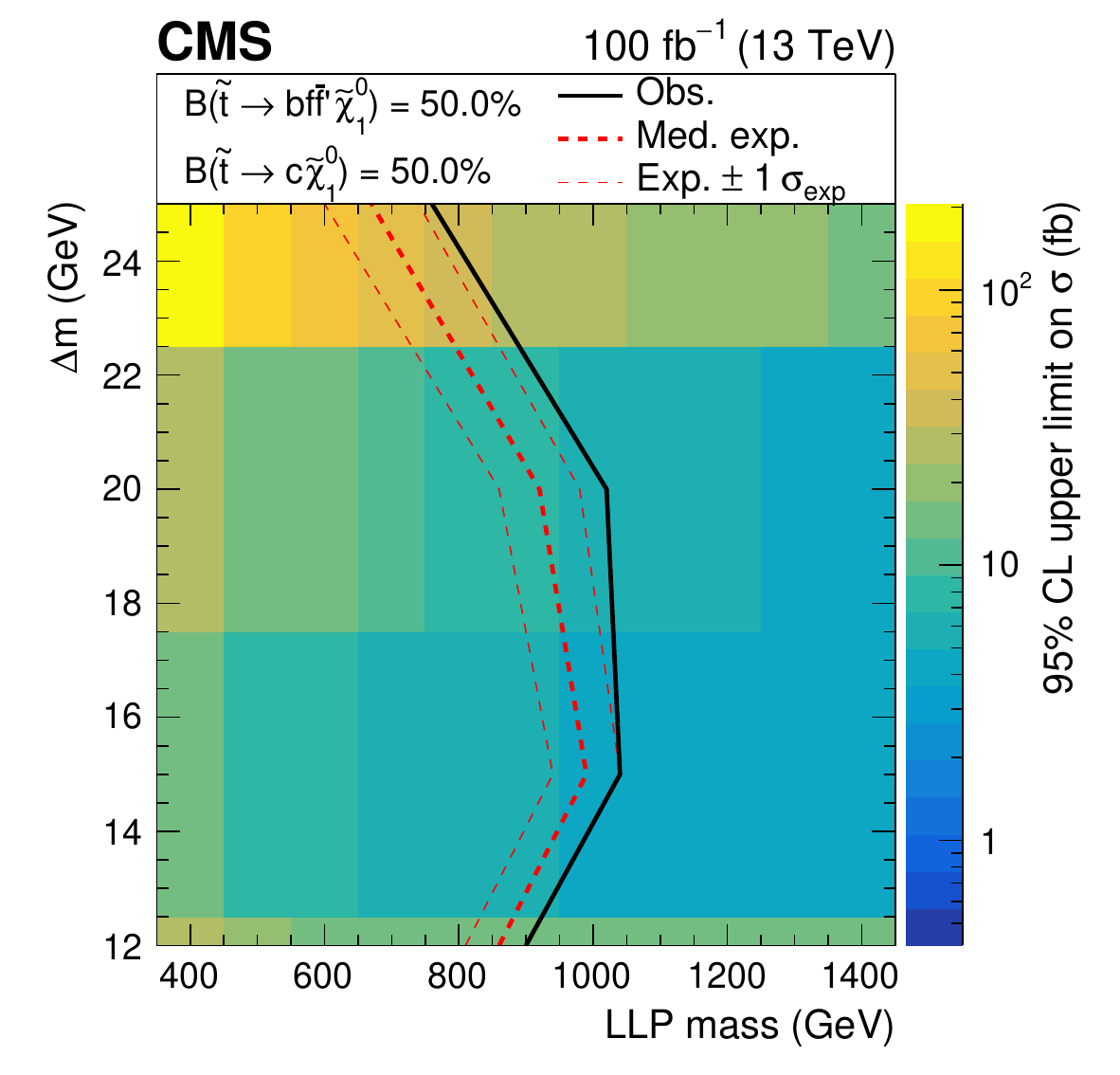}
    \includegraphics[width=0.45\textwidth]{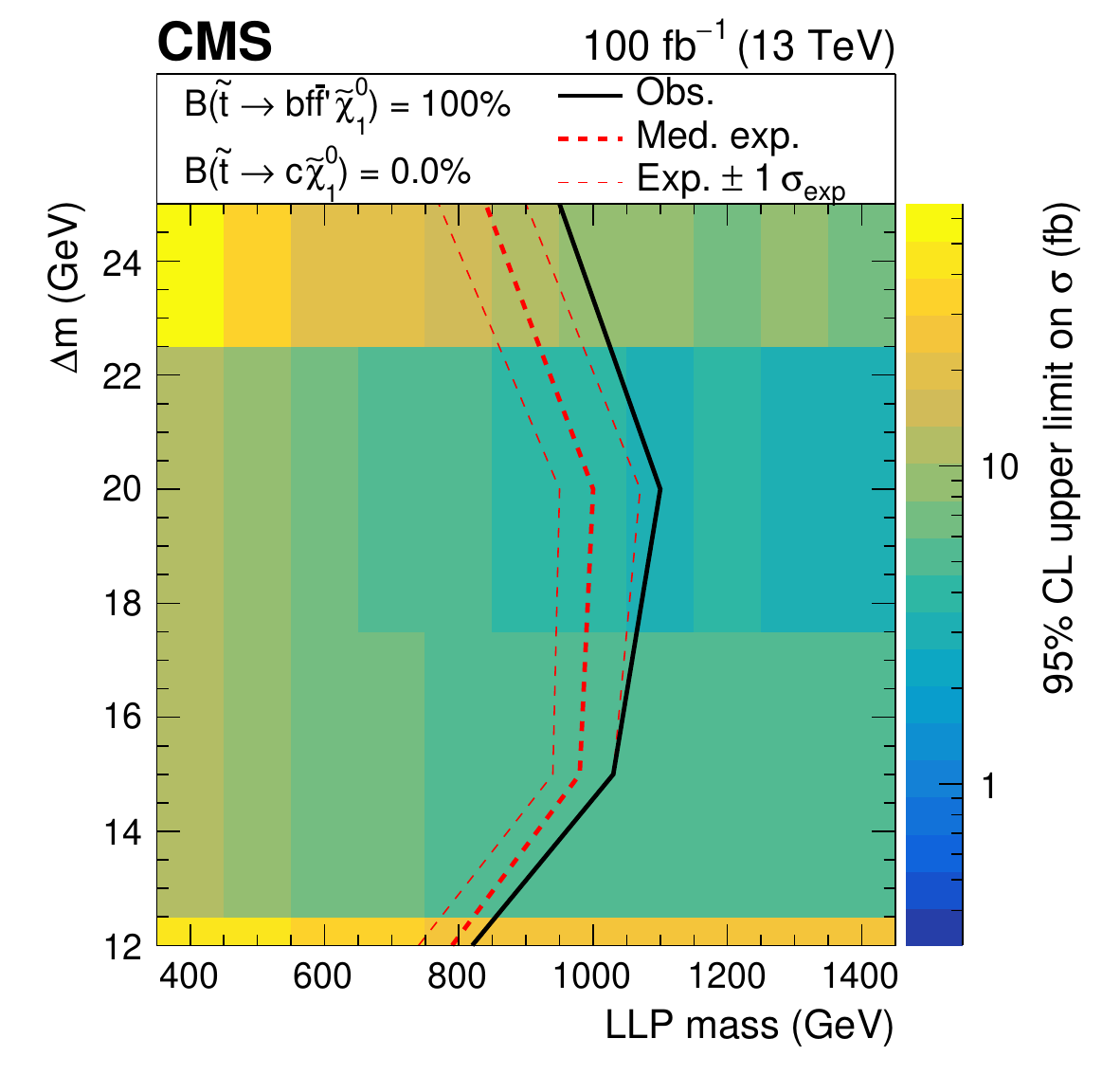}
    \caption{Observed 95\% \CL upper limits on the \PSQt production cross section, as functions of \mstop and \dm, for $\mathcal{B}(\PSQt \to \PQb \Pf\overline{\Pf}' \PSGczDo)$ of 10\% (upper left), 50\% (upper right), and 100\% (lower). The observed (solid black) and expected (dashed red) exclusion curves are overlaid on the plots. The search excludes the region to the left of the exclusion curves. For $\mathcal{B}(\PSQt \to \PQb \Pf\overline{\Pf}' \PSGczDo)=10\%$ (upper left), the exclusion curves terminate at $\dm = 20\GeV$, as no masses can be excluded at larger mass splittings because of reduced sensitivity in this region.}
    \label{fig:limit2d_stop}
\end{figure}

\begin{figure}[htbp]
    \centering
    \includegraphics[width=0.45\textwidth]{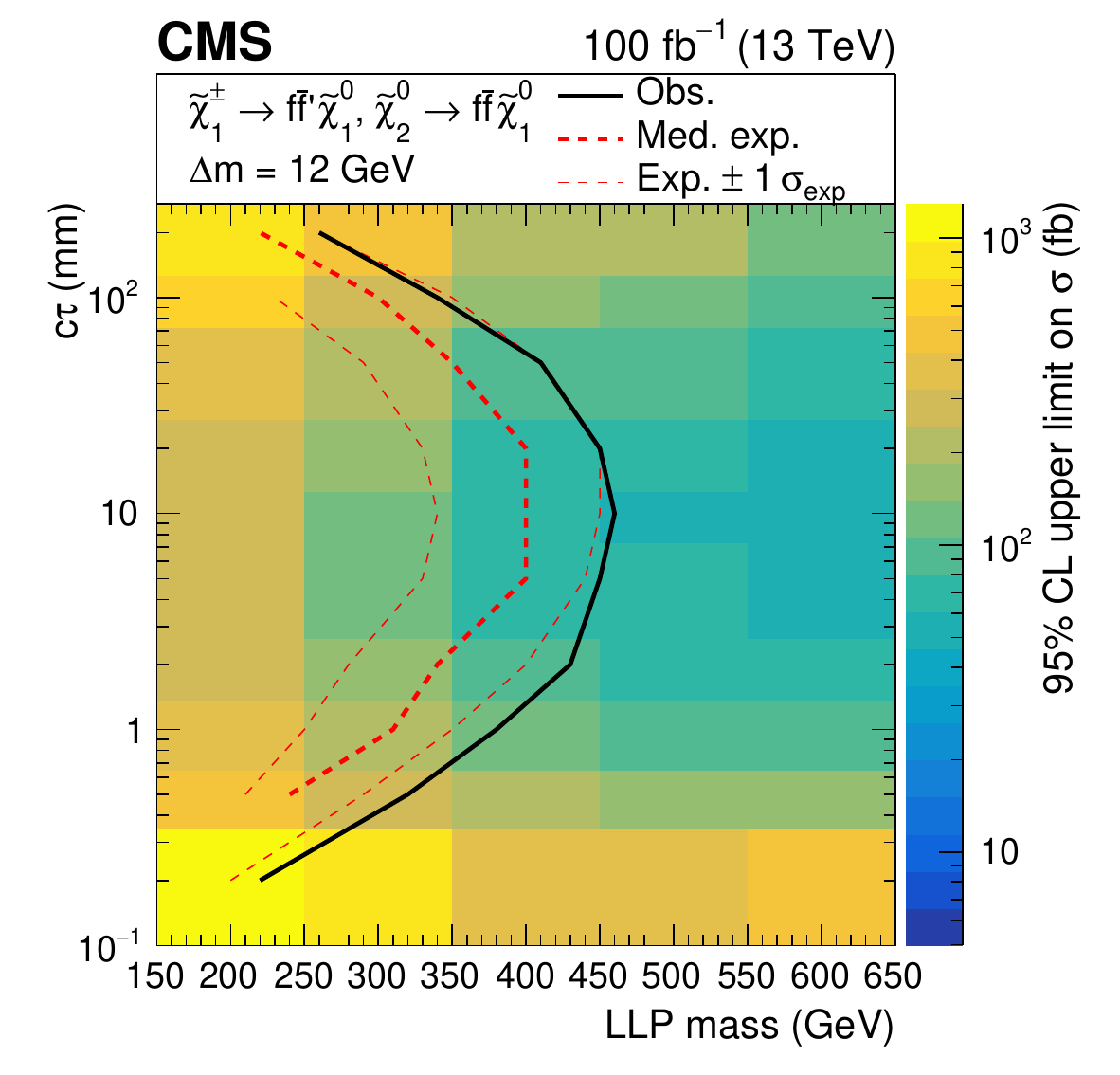}
    \includegraphics[width=0.45\textwidth]{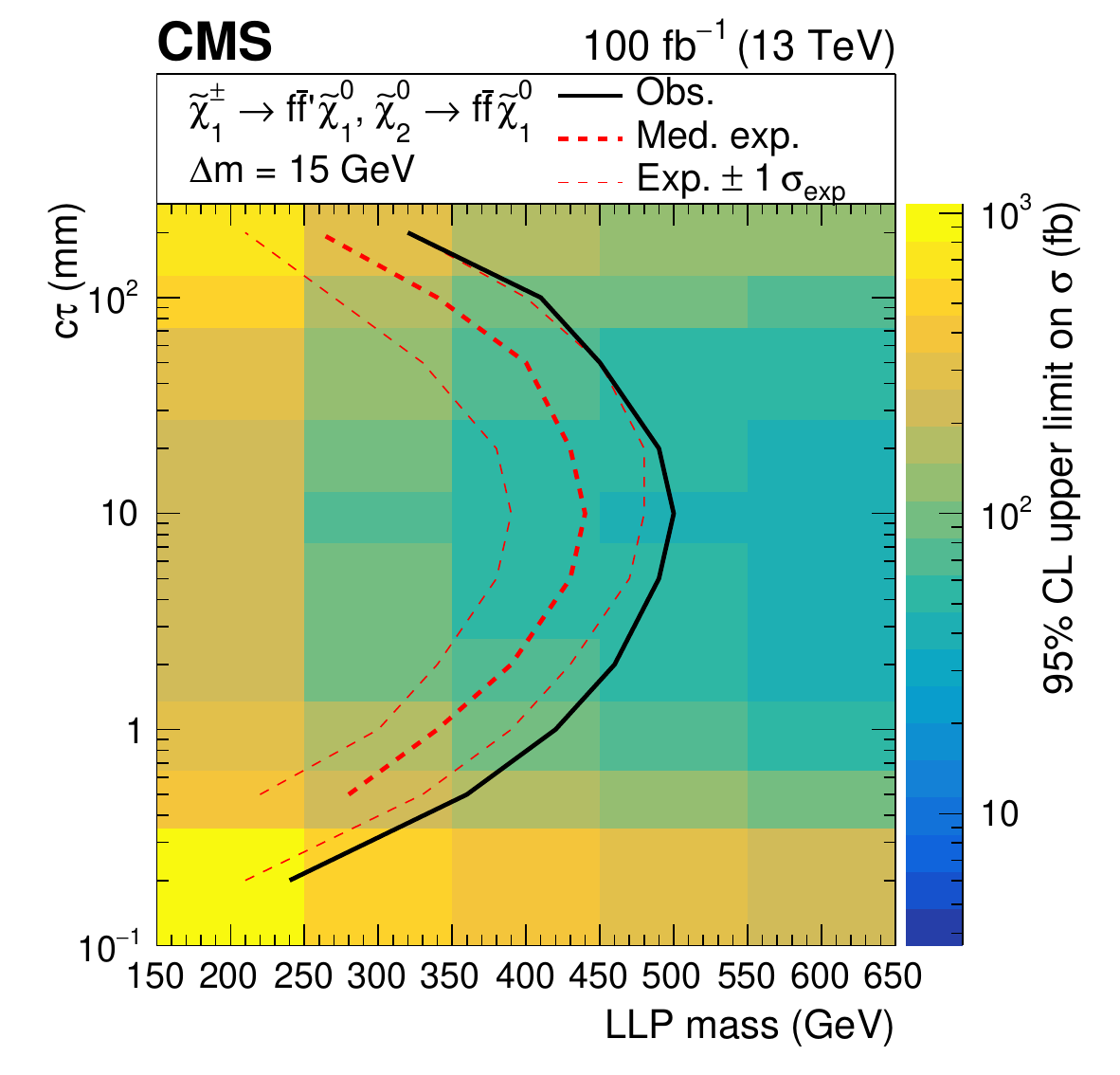}
    \includegraphics[width=0.45\textwidth]{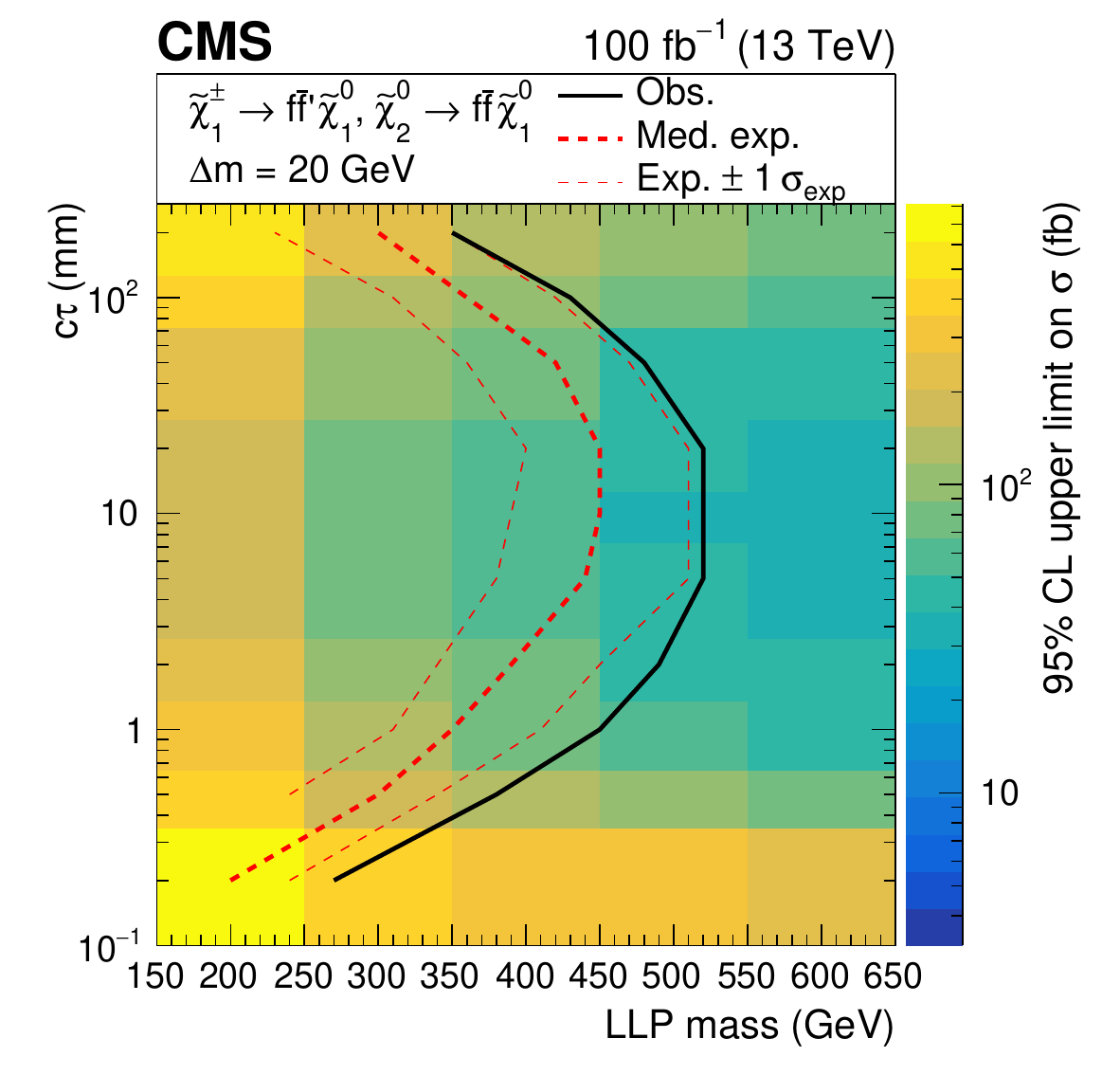}
    \includegraphics[width=0.45\textwidth]{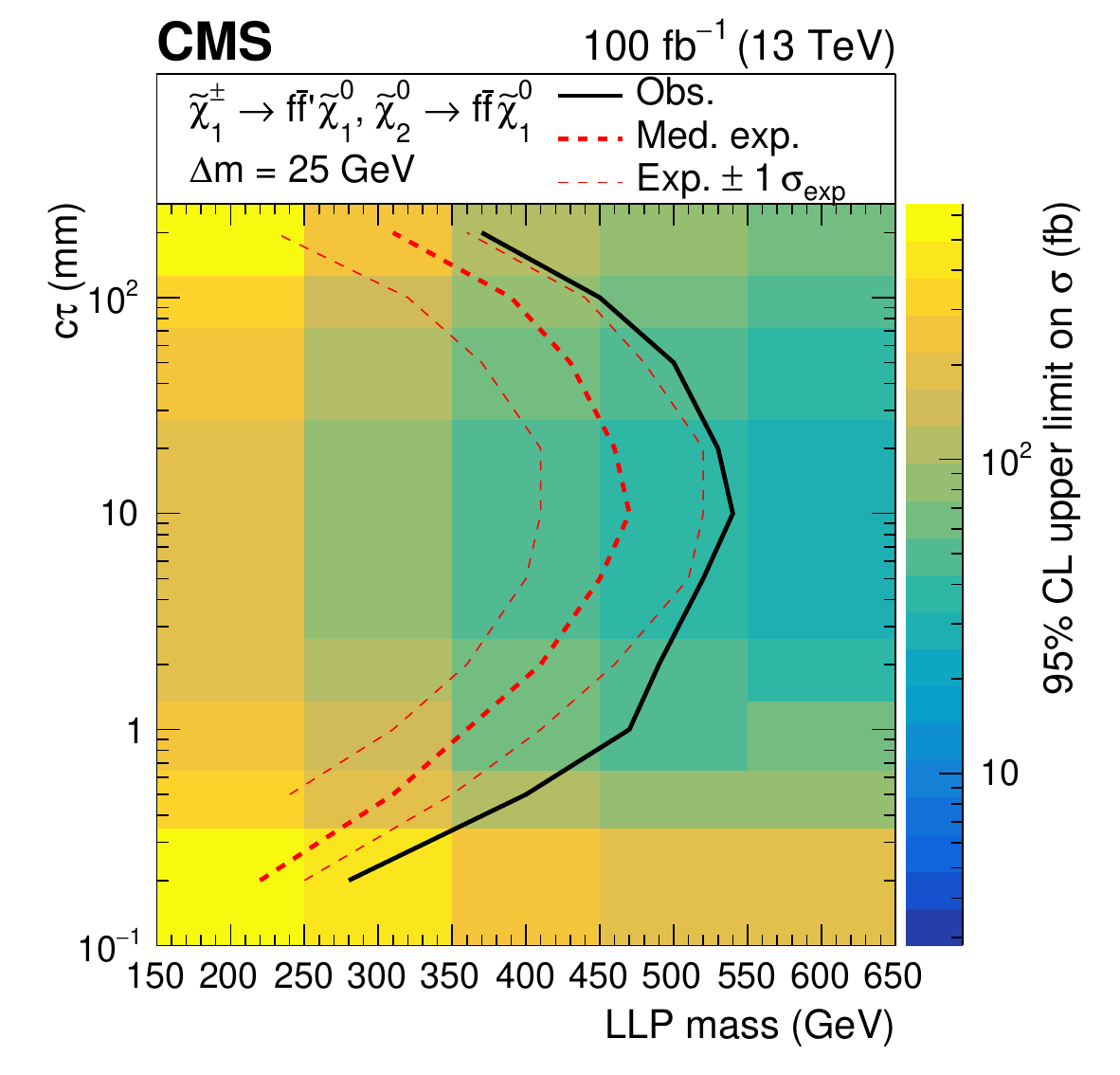}
    \caption{Observed 95\% \CL upper limits on the production cross section for the bino-wino NLSP model, as functions of \mllp and \ctau, for \dm of 12\GeV (upper left), 15\GeV (upper right), 20\GeV (lower left), and 25\GeV (lower right). The observed (solid black) and expected (dashed red) exclusion curves are overlaid on the plots. The search excludes the region to the left of the exclusion curves.}
    \label{fig:limit2d_c1n2}
\end{figure}

Figures~\ref{fig:limit2d_stop} and~\ref{fig:limit2d_c1n2} present the upper limits and exclusion regions for the \PSQt NLSP model and the bino-wino NLSP model, respectively. For the \PSQt NLSP model, the limits are shown as functions of \dm and \mllp under different assumptions of \br. For the bino-wino NLSP model, they are displayed as functions of the \mllp and \ctau, with different \dm scenarios shown in separate plots. The observed upper limits are lower than the expected ones due to differences between the observed and predicted yields in regions B and D of the tight signal plane.

The search sensitivity depends strongly on the \ctau. The optimal sensitivity is achieved for moderate values of \ctau around 10\mm. Sensitivity degrades at low \ctau because of the displacement requirements in track and vertex reconstruction, and at high \ctau because of reduced reconstruction and selection efficiency for highly displaced tracks. Sensitivity also improves with increasing \dm.

In the \PSQt NLSP model, where \ctau depends on \br, \dm, and \mllp, the region of highest sensitivity shifts with \br because of the change of \ctau. Specifically, the value of \dm that corresponds to a \ctau on the order of $\mathcal{O}(10\mm)$ increases with increasing \br. In contrast, for the bino-wino NLSP model, where the parameters are uncorrelated, the sensitivity is maximized for large \dm and moderate \ctau.

For the \PSQt NLSP model, the search excludes \PSQt with masses less than 400--1100\GeV, corresponding to theoretical production cross sections of 2150--3\unit{fb}, depending on the \dm and \br. This parameter space has not been explored by previous searches. For the bino-wino NLSP model, the search excludes NLSPs with masses below 220--550\GeV, corresponding to theoretical production cross sections of 1268--30\unit{fb}, depending on the \dm and \ctau.

\section{Summary}
\label{sec:summary}

A search for long-lived particles in signatures with displaced vertices with low-momentum tracks, missing transverse momentum, and an initial-state radiation jet has been presented. Proton-proton collision data at a center-of-mass energy of 13\TeV collected by the CMS experiment at the CERN LHC, with a total integrated luminosity of 100\fbinv, are used in the search. Compared with the previous CMS and ATLAS searches using displaced vertices, this search targets vertices with tracks of significantly lower momenta. This search adopts specific supersymmetric (SUSY) coannihilation scenarios as benchmark signal models, characterized by a long-lived next-to-lightest SUSY particle (NLSP) with a mass difference of less than 25\GeV relative to the lightest SUSY particle, assumed to be a bino-like neutralino. In the top squark (\PSQt) NLSP model, the NLSP is a \PSQt, while in the bino-wino NLSP scenario, the mass-degenerate NLSPs are a wino-like long-lived neutralino and a short-lived chargino.

This search reconstructs displaced vertices using a customized algorithm based on the inclusive vertex finder~\cite{CMS:2017wtu}. In addition, the background estimation method using transfer factors allows for targeting multiple signal regions, thus enhancing the search sensitivity.

The search shows good overall agreement between the background predictions and observed event yields across most of the signal regions. The search excludes \PSQt masses less than 400--1100\GeV and wino-like neutralinos with masses less than 220--550\GeV, depending on the signal parameters. This search is the first at the LHC to demonstrate sensitivity to long-lived particles in compressed-spectrum scenarios using displaced-vertex signatures. It sets the most stringent upper limits to date for the \PSQt and bino-wino NLSP signal models.

\begin{acknowledgments}
We congratulate our colleagues in the CERN accelerator departments for the excellent performance of the LHC and thank the technical and administrative staffs at CERN and at other CMS institutes for their contributions to the success of the CMS effort. In addition, we gratefully acknowledge the computing centers and personnel of the Worldwide LHC Computing Grid and other centers for delivering so effectively the computing infrastructure essential to our analyses. Finally, we acknowledge the enduring support for the construction and operation of the LHC, the CMS detector, and the supporting computing infrastructure provided by the following funding agencies: SC (Armenia), BMBWF and FWF (Austria); FNRS and FWO (Belgium); CNPq, CAPES, FAPERJ, FAPERGS, and FAPESP (Brazil); MES and BNSF (Bulgaria); CERN; CAS, MoST, and NSFC (China); MINCIENCIAS (Colombia); MSES and CSF (Croatia); RIF (Cyprus); SENESCYT (Ecuador); ERC PRG, TARISTU24-TK10 and MoER TK202 (Estonia); Academy of Finland, MEC, and HIP (Finland); CEA and CNRS/IN2P3 (France); SRNSF (Georgia); BMFTR, DFG, and HGF (Germany); GSRI (Greece); NKFIH (Hungary); DAE and DST (India); IPM (Iran); SFI (Ireland); INFN (Italy); MSIT and NRF (Republic of Korea); MES (Latvia); LMTLT (Lithuania); MOE and UM (Malaysia); BUAP, CINVESTAV, CONACYT, LNS, SEP, and UASLP-FAI (Mexico); MOS (Montenegro); MBIE (New Zealand); PAEC (Pakistan); MES, NSC, and NAWA (Poland); FCT (Portugal); MESTD (Serbia); MICIU/AEI and PCTI (Spain); MOSTR (Sri Lanka); Swiss Funding Agencies (Switzerland); MST (Taipei); MHESI (Thailand); TUBITAK and TENMAK (T\"{u}rkiye); NASU (Ukraine); STFC (United Kingdom); DOE and NSF (USA).

\hyphenation{Rachada-pisek} Individuals have received support from the Marie-Curie program and the European Research Council and Horizon 2020 Grant, contract Nos.\ 675440, 724704, 752730, 758316, 765710, 824093, 101115353, 101002207, 101001205, and COST Action CA16108 (European Union); the Leventis Foundation; the Alfred P.\ Sloan Foundation; the Alexander von Humboldt Foundation; the Science Committee, project no. 22rl-037 (Armenia); the Fonds pour la Formation \`a la Recherche dans l'Industrie et dans l'Agriculture (FRIA) and Fonds voor Wetenschappelijk Onderzoek contract No. 1228724N (Belgium); the Beijing Municipal Science \& Technology Commission, No. Z191100007219010, the Fundamental Research Funds for the Central Universities, the Ministry of Science and Technology of China under Grant No. 2023YFA1605804, the Natural Science Foundation of China under Grant No. 12061141002, 12535004, and USTC Research Funds of the Double First-Class Initiative No.\ YD2030002017 (China); the Ministry of Education, Youth and Sports (MEYS) of the Czech Republic; the Shota Rustaveli National Science Foundation, grant FR-22-985 (Georgia); the Deutsche Forschungsgemeinschaft (DFG), among others, under Germany's Excellence Strategy -- EXC 2121 ``Quantum Universe" -- 390833306, and under project number 400140256 - GRK2497; the Hellenic Foundation for Research and Innovation (HFRI), Project Number 2288 (Greece); the Hungarian Academy of Sciences, the New National Excellence Program - \'UNKP, the NKFIH research grants K 131991, K 133046, K 138136, K 143460, K 143477, K 146913, K 146914, K 147048, 2020-2.2.1-ED-2021-00181, TKP2021-NKTA-64, and 2021-4.1.2-NEMZ\_KI-2024-00036 (Hungary); the Council of Science and Industrial Research, India; ICSC -- National Research Center for High Performance Computing, Big Data and Quantum Computing, FAIR -- Future Artificial Intelligence Research, and CUP I53D23001070006 (Mission 4 Component 1), funded by the NextGenerationEU program (Italy); the Latvian Council of Science; the Ministry of Education and Science, project no. 2022/WK/14, and the National Science Center, contracts Opus 2021/41/B/ST2/01369, 2021/43/B/ST2/01552, 2023/49/B/ST2/03273, and the NAWA contract BPN/PPO/2021/1/00011 (Poland); the Funda\c{c}\~ao para a Ci\^encia e a Tecnologia, grant CEECIND/01334/2018 (Portugal); the National Priorities Research Program by Qatar National Research Fund;h MICIU/AEI/10.13039/501100011033, ERDF/EU, "European Union NextGenerationEU/PRTR", and Programa Severo Ochoa del Principado de Asturias (Spain); the Chulalongkorn Academic into Its 2nd Century Project Advancement Project, the National Science, Research and Innovation Fund program IND\_FF\_68\_369\_2300\_097, and the Program Management Unit for Human Resources \& Institutional Development, Research and Innovation, grant B39G680009 (Thailand); the Kavli Foundation; the Nvidia Corporation; the SuperMicro Corporation; the Welch Foundation, contract C-1845; and the Weston Havens Foundation (USA).    

\end{acknowledgments}

\bibliography{auto_generated}
\cleardoublepage \appendix\section{The CMS Collaboration \label{app:collab}}\begin{sloppypar}\hyphenpenalty=5000\widowpenalty=500\clubpenalty=5000\cmsinstitute{Yerevan Physics Institute, Yerevan, Armenia}
{\tolerance=6000
A.~Hayrapetyan, V.~Makarenko\cmsorcid{0000-0002-8406-8605}, A.~Tumasyan\cmsAuthorMark{1}\cmsorcid{0009-0000-0684-6742}
\par}
\cmsinstitute{Institut f\"{u}r Hochenergiephysik, Vienna, Austria}
{\tolerance=6000
W.~Adam\cmsorcid{0000-0001-9099-4341}, J.W.~Andrejkovic, L.~Benato\cmsorcid{0000-0001-5135-7489}, T.~Bergauer\cmsorcid{0000-0002-5786-0293}, M.~Dragicevic\cmsorcid{0000-0003-1967-6783}, C.~Giordano, A.K.~Guven, P.S.~Hussain\cmsorcid{0000-0002-4825-5278}, M.~Jeitler\cmsAuthorMark{2}\cmsorcid{0000-0002-5141-9560}, N.~Krammer\cmsorcid{0000-0002-0548-0985}, A.~Li\cmsorcid{0000-0002-4547-116X}, D.~Liko\cmsorcid{0000-0002-3380-473X}, M.~Matthewman, I.~Mikulec\cmsorcid{0000-0003-0385-2746}, J.~Schieck\cmsAuthorMark{2}\cmsorcid{0000-0002-1058-8093}, R.~Sch\"{o}fbeck\cmsAuthorMark{2}\cmsorcid{0000-0002-2332-8784}, D.~Schwarz\cmsorcid{0000-0002-3821-7331}, M.~Shooshtari\cmsorcid{0009-0004-8882-4887}, M.~Sonawane\cmsorcid{0000-0003-0510-7010}, W.~Waltenberger\cmsorcid{0000-0002-6215-7228}, C.-E.~Wulz\cmsAuthorMark{2}\cmsorcid{0000-0001-9226-5812}
% ...
\par}
\vskip\cmsinstskip
\dag:~Deceased\\
$^{1}$Also at Yerevan State University, Yerevan, Armenia\\
$^{2}$Also at TU Wien, Vienna, Austria\\
% ...
\end{sloppypar}
\end{document}